\documentclass[%
    aps,
    prd,
    12pt,
    nofootinbib,
    eprint,
    amsmath,
    amssymb
]{revtex4-2}

\usepackage{tcolorbox}

\usepackage{graphicx}
\usepackage[%
    colorlinks=true,
    allcolors=blue
]{hyperref}
\usepackage[%
    print-unity-mantissa=false,
    range-phrase=-
]{siunitx}

\usepackage{cleveref}

\newcommand{\Ibar}{{\mathcal I} \kern -0.5em\raise 0.25ex \hbox{--}}

\usepackage{geometry}
\usepackage{graphicx}
\usepackage{xcolor}
\definecolor{DarkBlue}{rgb}{0,0,0.4}
\usepackage{natbib}

\usepackage{derivative}

\newcommand{\tg}{\ensuremath{\text{g}}}

\def\spose#1{\hbox to 0pt{#1\hss}}

\def\lta{\mathrel{\spose{\lower 3pt\hbox{$\mathchar"218$}}
     \raise 2.0pt\hbox{$\mathchar"13C$}}}
\def\gta{\mathrel{\spose{\lower 3pt\hbox{$\mathchar"218$}}
     \raise 2.0pt\hbox{$\mathchar"13E$}}}

\newcommand{\cf}{\textsl{cf.~}}
\newcommand{\eg}{\textsl{e.g.}}

\def\beq{\begin{equation}}
\def\eeq{\end{equation}}
\def\bea{\begin{eqnarray}}
\def\eea{\end{eqnarray}}

\def\n{{\rm n}}

\def\s{{\rm s}}

\def\fint{\int_{\cal M} {\rm d}^4 x \sqrt{- g}}

\newcommand{\D}{{\cal D}}
\newcommand{\Ab}{{\bar{A}}}
\newcommand{\Bb}{{\bar{B}}}
\newcommand{\Cb}{{\bar{C}}}
\newcommand{\Db}{{\bar{D}}}
\newcommand{\Eb}{{\bar{E}}}
\newcommand{\Fb}{{\bar{F}}}

\def\bu{\bar{u}}

\def\xa{X^A}

\def\xd{X^D}

\def\bxa{\bar{X}^\Ab}

\def\bxd{\bar{X}^\Db}

\def\paa{\Psi^A_a}
\def\pab{\Psi^A_b}
\def\pac{\Psi^A_c}

\def\pbb{\Psi^B_b}

\def\pcc{\Psi^C_c}

\def\pdd{\Psi^D_d}
\def\pda{\Psi^D_a}
\def\pdb{\Psi^D_b}

\def\peb{\Psi^E_b}

\def\ped{\Psi^E_d}

\def\bpaa{\bar{\Psi}^\Ab_a}
\def\bpab{\bar{\Psi}^\Ab_b}
\def\bpac{\bar{\Psi}^\Ab_c}
\def\bpbb{\bar{\Psi}^\Bb_b}
\def\bpba{\bar{\Psi}^\Bb_a}

\def\bpcc{\bar{\Psi}^\Cb_c}

\def\bpdd{\bar{\Psi}^\Db_d}
\def\bpda{\bar{\Psi}^\Db_a}

\def\bpeb{\bar{\Psi}^\Eb_b}

\def\Bn{{\cal B}^\n}

\def\Bs{{\cal B}^\s}
\def\Asn{{\cal A}^{\s \n}}

\def\gab{g^{A B}}

\def\gba{g^{B A}}

\def\gde{g^{D E}}

\def\bgab{\bar{g}^{\Ab \Bb}}

\def\bgba{\bar{g}^{\Bb \Ab}}

\def\bgde{\bar{g}^{\Db \Eb}}

\def\hgab{\hat{g}^{A \Bb}}
\def\hgba{\hat{g}^{\Bb A}}

\def\ginvab{g_{A B}}
\def\ginvac{g_{A C}}

\def\ginvbd{g_{B D}}

\def\bginvab{\bar{g}_{\Ab \Bb}}

\def\hginvab{\hat{g}_{A \Bb}}

\def\dgab{{\cal L}_u g^{A B}}

\def\dgcd{{\cal L}_u g^{C D}}
\def\dgde{{\cal L}_u g^{D E}}
\def\dbugab{{\cal L}_{\bu} g^{A B}}

\def\bdgab{{\cal L}_u \bar{g}^{\Ab \Bb}}
\def\dbubgab{{\cal L}_{\bu} \bar{g}^{\Ab \Bb}}

\def\dbungab{{\cal L}_u \bar{g}^{\Ab \Bb}}

\def\hdgab{{\cal L}_u \hat{g}^{A \Bb}}
\def\dhubgab{{\cal L}_{\bu} \hat{g}^{A \Bb}}

\begin{document}

\title{Action based approach to dissipative relativistic fluid systems}

\author{G. L. Comer\footnote{Greg Comer sadly passed away as this paper was in the final stages of completion.}, N. Andersson$^{1}$,  
T. Celora$^{2}$ and I. Hawke$^{1}$}

\affiliation{$^1$Mathematical Sciences and STAG Research Centre, University of Southampton, Southampton SO17 1BJ, United Kingdom}

\affiliation{$^2$ Institute of Space Sciences (ICE, CSIC), Campus UAB, Carrer de Magrans, 08193 Barcelona, Spain}

\begin{abstract}
We develop an action principle for a relativistic two-fluid system with dissipation. The specific constituents of the model---which serves as a proof of principle---are particles and entropy. 
The linchpin of the action is the assertion that a given flux is dissipative if its covariant divergence is non-zero. For our model, the particle flux is taken to be conservative while the entropy flux is dissipative. This allows for a ``top-down'' approach where the general question is geometric: How can dissipative flux vector fields (non-zero divergences) be embedded in curved spacetime solutions to the Einstein field equations? The answer lies in the action principle, which produces evolution equations whose solutions are precisely the embeddings we seek. Previous work has shown that new terms (the proper time derivative of matter space ``metrics'') must be included in the Lagrangian in order to produce equations of motion with terms representing bulk and shear viscosity. In addition to including these terms we show that further terms---interpreted as velocities---can be included. The new action-based model recovers known relativistic formulations of the Cattaneo equation, which results in causal heat propagation. We further advance our understanding by exploring the single-fluid limit by locking the entropy four-velocity  to that of the matter component. This reduces the system to a single field equation along with a constraint equation. We show that this constraint leads to a dynamical extension of the standard Tolman red-shift condition. Finally, we provide three example actions (of increasing complexity) which demonstrate that the model is able to reproduce (in the single-fluid limit) the anticipated terms from the relativistic Navier-Stokes equations. In the general case, the action based approach allows for a much richer structure, which may be relevant for realistic models of
non-linear
dissipative relativistic fluid systems.
\end{abstract}

\maketitle

\section{Introduction}

\noindent
Thermodynamical studies of general relativistic many-body systems become problematic if they adhere strictly to the intuition and reasoning 
derived from systems where gravitational effects are ignored. The problem is that, even though the Einstein field equations are based on the equivalence 
principle and the existence of locally flat metrics, there are no global symmetries (such as time- and 
space-rotational/translational invariance) one may invoke for generic spacetimes. Consequently, there are no generally accepted notions of Noether-conserved total energy and 
(linear/angular) momentum (\cf Wheeler's ``nowhere to stand'' discussion in \cite{mtw73}).
This is bad news for textbook-style thermodynamics since 
energy and (linear/angular) momentum conservation for the total system are typically taken as given (see, for example, 
\cite{landau1980:_statistical}). It is commonly assumed that the total system can be broken up into subsystems with their own energies and momenta, 
such that the simple sum of the individual energies and momenta are equal to that of the total system. However, this is not true in general relativity. 

The basic issue, from a general relativity perspective, is that the laws of thermodynamics are commonly  established from ``local 
enough''---observations/experiments (e.g. gravitational tidal forces are negligible, $\Delta R^{a b c d} \sim 1/L^2_{\mathrm{tf}}$, where $R^{a b c d}$ is the Riemann 
tensor). For example, for the Earth with radius $R_E$ we have $L_{\mathrm{tf}} \sim R_E$ which is much greater than the typical size 
of reliable thermal systems like engines, refrigerators, power plants, and so on. Local enough also means that one may introduce effective time-, 
and space-translational/rotational-invariances leading to a Noether conserved total energy and translational/rotational momenta. 

The typical use of the word ``universe'' in classical thermodynamics carries with it an implicit assumption that local symmetries exist globally. 
Modern cosmology demonstrates that this assumption is not valid. Landau and Lifschitz \cite{landau1980:_statistical} also 
call into question how the Universe is inserted into thermodynamic analyses, but from the statistical mechanics perspective the universe 
itself is expanding and hence cannot be in an equilibrium state. The  laws of thermodynamics are therefore statements hinging on the existence 
of these spacetime symmetries and the overall size of the spacetime volume occupied during any system evolution. In essence, a thermodynamic 
description of a general relativistic system has to deal with the global, non-screenable reach of gravity, the lack of flat-space symmetries for 
generic spacetimes and the tidal nature of its interaction with matter and energy.      

The notion of equilibrium is crucial to traditional thermodynamic studies because it sets up an endpoint to which a system is driven as it 
evolves. The presence of gravity is known to impact on the usual statements of chemical, mechanical and thermodynamical equilibrium: pressure will not 
necessarily be the same throughout the system for mechanical equilibrium and the chemical potential and temperature change due to the so-called Tolman red-shift \cite{tolman34:_book}. For example, a huge lake can have effectively net zero momentum throughout, implying it is at 
rest and in mechanical equilibrium, but one only needs to dive in and swim towards the bottom to feel the change in pressure on the ears. 
Less evident is the gravitational effect on the lake's temperature because of the red-shift of energy; one needs more precise 
measurements than ears can provide, such as the Pound-Rebka experiment \cite{Pound59:_red-shift}. Alternatively, we need to abandon the lake  and examine light emerging from the 
vicinity of strongly gravitating objects like neutron stars and black holes. We may even consider the much bigger system of the visible piece of the Universe 
and the effect its expansion has on the cosmic background blackbody radiation \cite{xia24:_generaltolman}. 

Accepting these massive potholes in the road ahead, we explore the problem of heat/entropy through a variational approach which assumes that the natural starting 
point is a system with two fluxes---one for the particles and another for the entropy (the basic definition of heat requires it to be a flux). The 
associated action implies no restrictions on the global nature of spacetime. Instead, the main assumption is that there exists a Lagrangian that is a functional of all scalars that can be formed out of the two fluxes. As previously discussed in \cite{Andersson15:_dissfl_act} and \cite{andersson24:_sinffldiss}, dissipation enters the formalism when the dual three-form to the entropy flux is no longer closed 
which immediately implies that there is a non-zero entropy creation rate. The action is also non-linear in the sense that it is not based on an 
expansion away from some prescribed equilibrium. The fact that the model is nonlinear makes it distinct from most other currently available approaches to the problem. However, as we shall see, this means that one has to carefully consider the notion of equilibrium for a generic spacetime. Intimately connected with this, it also seems that defining equilibrium as the maximization of the entropy is problematic. Entropy is a flux and in order to create a scalar to make a maximization argument we have
contract it with worldlines associated with a frame of reference erected in the relevant spacetime region. In effect, the question of whether or not the entropy is maximized is frame dependent. A minimal requirement for any notion of equilibrium is that the entropy flux evolves to become comoving with the particle flow, with no residual heat flux.

By exploring the equations of motion that result from a set of proposed actions 
\cite{andersson21:_livrev,Andersson15:_dissfl_act,andersson24:_ent1diss}, we develop a ``top-down'' approach to general relativistic 
many-particle thermodynamics. The action principle provides an equation of motion for the particles and an 
independent equation for the entropy. Their relative fluxes are dynamically independent, 
meaning that their respective trajectories in spacetime will not be parallel. We  use both sets of equations of motion to explore consequences of the 
anticipated ``locking'' together of the two fluids (leading to them  having the same four-velocity) due to dissipation. This typically leads to a Cattaneo-type equation \cite{Cattaneo48:_cateq} for each of the models we consider. This equation governs the entropy drift relative to the particles. An immediate implication is the emergence of constraint on the dynamics that, in the simplest case, leads to the Tolman temperature red-shift phenomenon. 

In developing the variational model, it is important to note that the second law of thermodynamics does not follow from generic forms of the action; rather, it must be imposed 
from ``outside'' the chosen framework. Again, the problem with entropy is that it is a flow of heat which implies a
frame-dependent notion of an entropy scalar.  The process of imposing the second law from outside then means finding specific actions which 
make the divergence of the entropy flux positive-definite. For the particular models examined here, the entropy creation rate in the non-linear regime 
is---perhaps not surprisingly---not amenable to being written in a positive-definite form. We can make progress, however, by writing the equations of motion in the particle 
frame and  expanding them in terms of the relative entropy drift. In the limit where this drift is sufficiently small one can demonstrate that the 
second law kicks in and that the
global  total entropy in an extended region of spacetime grows. Moreover, in this limit one can show that the action-based model reduces in a natural way to the standard relativistic Navier-Stokes equations. This is an important proof-of-principle demonstration.

The paper is laid out as follows: 
In Sec.~\ref{zerodissfluids}, we consider non-dissipative systems and explore the single-fluid limit. In Sec.~\ref{flact}, the generic action is written 
down and a variation with respect to the field variables (particle and entropy flux, the spacetime metric, and the entrainment parameter which is 
included throughout) is given. Abstract three-dimensional ``matter'' spaces are introduced so that the fluxes can be 
reformulated in such a way that the action principle becomes viable (see \cite{comer93:_hamil_multi_con,comer94:_hamil_sf} for an earlier 
application of this process). All the degrees of freedom are introduced there, including an additional set not included in 
previous work, \cite{Andersson15:_dissfl_act} and \cite{andersson24:_ent1diss}. Sec.~\ref{lagdis} takes the degrees of freedom introduced in 
the previous section and builds their Lagrangian displacements. Following this, Sec.~\ref{fldeqs} takes the jump from using matter space degrees 
of freedom to their related spacetime version. It is shown how the dissipative channels enter the action principle in a non-linear way. In 
Sec.~\ref{spevar}, three specific, in terms of degrees of freedom, actions are investigated. The equations of motion, stress-energy-momentum 
tensor, and entropy creation rate for each action are derived. In Sec.~\ref{expan} the single-fluid limit is explored by writing the equations with 
respect to the particle frame and then expanding to linear order in the entropy drift. Specifically, we write down a set of single fluid equations which 
manifestly satisfy the second law. Concluding remarks are offered in Sec.~\ref{conclu}. Additional details are relegated to two Appendices. In Appendix \ref{appen} we present some 
formulas that complete the Lagrangian variations presented in Sec.~\ref{lagdis}. Finally, in Appendix \ref{app:Tolman} we provide the derivation/generalization of the Tolman effect for more complicated spacetimes.

Many of the ideas and derivations presented here are already discussed in detail in \cite{andersson24:_ent1diss}. This means that some results will just be quoted, with the understanding that the interested reader can 
explore \cite{andersson24:_ent1diss} for details. We use ``MTW'' \cite{mtw73} conventions throughout the paper.

\section{A non-dissipative two-fluid system}
\label{zerodissfluids}

\noindent
In order to set the stage for the discussion of dissipative systems, let us  consider a non-dissipative two-fluid problem. The natural model to explore is that of matter and heat, with the latter represented by an entropy flux (in turn treated as a ``fluid'', following \cite{andersson21:_livrev}). In essence, the problem involves a particle flux $n^a$, an entropy flux 
$s^a$, and the spacetime metric $g_{a b}$ (we use roman letters for spacetime indices throughout the analysis). The particle flux leads to a particle density $n$ obtained from $n^2 = - g_{a b} n^a n^b$, and similarly, the 
entropy density $s$ is obtained from $s^2 = - g_{a b} s^a s^b$. The two-fluid system also allows for the presence of entrainment, encoded in the parameter  $x^2 = - g_{a b} n^a s^b$. It is well established that the entrainment between matter and entropy plays an important role in endowing the entropy component with an effective mass, a key ingredient for a causal relativistic model of heat flux \cite{andersson21:_livrev}. As we will soon see, entrainment results in a ``tilting'' of a fluid's (particle and thermal) momentum with 
respect to its flux; i.e.,~the momentum and flux are no longer aligned. Further, we denote the  unit four-velocity for the particle flux as $u^a$, 
with normalization $u_a u^a = - 1$ (in geometric units), while the corresponding unit four-velocity for the entropy flux is $\bu^a$, with 
normalization $\bu_a \bu^a = - 1$. The particle flux then becomes $n^a = n u^a$ and the entropy flux is $s^a = s \bu^a$. The particle density can 
 be obtained from $n = - u_a n^a$ and the entropy density from $s = - \bu_a s^a$. 

The entire system is controlled by a spacetime Lagrangian, denoted by $\Lambda$. For an isotropic system (which we assume), the Lagrangian depends on each of the scalar parameters $n^2$, $s^2$, 
and $x^2$. The quantity $- \Lambda$ is the closest thing to an equation of state that a multi-fluid system can have (see \cite{andersson21:_livrev}  for further discussion). The main difficulty in building an 
equation of state for a multi-fluid system is that the equation of state is typically expressed in term of an energy density depending on the relevant state parameters. However, 
energy is a frame-dependent quantity (see, for example, the straightforward argument in \cite{Landau69:_mech_book}) and in our problem there are two 
natural frames of reference one may consider. Picking one frame over the other essentially comes down to a choice based on 
convenience; for example, in the analysis of the field equations derived below we will find it advantageous to project tensors into the particle 
frame (analogous to the so-called Eckart frame in discussions of relativistic thermodynamics). 

Despite this conceptual complication, $\Lambda$ is unambiguous in its service as the Lagrangian for the fluids in the action principle. It is well established 
(see \cite{andersson21:_livrev} for details and references to the broader literature) that a variation of $\Lambda$ leads to two sets of field equations of the forms
\begin{subequations}
\begin{align}
       2 n^b \nabla_{[b} \mu_{a]} & = 0 \ , \quad \nabla_a n^a = 0 \ , \\
       2 s^b \nabla_{[b} \Theta_{a]} & = 0 \ , \quad \nabla_a s^a = 0 \ ,
\label{cons_eqs}
\end{align}
\end{subequations}
where the chemical potential and thermal momenta are given by, respectively,
\begin{subequations}
\begin{align}
  \mu_a & =  - \left(2 \frac{\partial \Lambda}{\partial n^2} n_a + \frac{\partial \Lambda}{\partial x^2} s_a\right) 
                    \equiv \Bn n_a +  \Asn s_a \ , 
                    \label{genmom}\\
  \Theta_a & =  - \left(2 \frac{\partial \Lambda}{\partial s^2} s_a + \frac{\partial \Lambda}{\partial x^2} n_a\right) 
                    \equiv \Bs s_a +  \Asn n_a \ . \label{genTmom}              
\end{align}
\end{subequations}
The generalized chemical potential is defined as $\mu_\tg = - \mu_a u^a$ while the generalized temperature is given by $T_\tg = - \Theta_a \bu^a$. We also have  
a generalized pressure $\Psi$ given by
\begin{align}
  \Psi = \Lambda - \mu_a n^a - \Theta_a s^a \ . 
  \label{genpress}
\end{align}

\subsection{The Single-fluid Limit}

\noindent
It is instructive to consider how the two-fluid system limits to the, likely more familiar, single-fluid situation. In order to obtain a  single-fluid system we can simply set $\bu^a = u^a$. This implies\footnote{Note that, if we had started from an action principle with the two 
constituents $n$ and $s$ and a single four-velocity $u^a$, then we would have $\Asn = 0$, see \cite{andersson21:_livrev}.} 
\begin{subequations}
\begin{align}
      \mu_a & = \left(\Bn n + \Asn s\right) u_a \ , \quad \mu_\tg = \Bn n + \Asn s \ , 
      \label{singflmug} \\
      \Theta_a & = \left(\Bs s + \Asn n\right) u_a \ , \quad T_\tg = \Bs s + \Asn n 
      \label{singflTg} \ .
\end{align}
\end{subequations}
In turn, the equations of motion become
\begin{subequations}
\begin{align}
      \perp_a^b n \nabla_b \mu_\tg + \mu_\tg n u^b \nabla_b u_a & = 0 \ , \quad u^a \nabla_a n + n \nabla_a u^a = 0 \ , \\
      \perp_a^b s \nabla_b T_\tg + T_\tg s u^b \nabla_b u_a & = 0 \ , \quad u^a \nabla_a s + s \nabla_a u^a = 0 \ ,
      \label{entcons}
\end{align}
\end{subequations}
where $\perp^a_b = \delta^a_b + u^a u_b$. Noting that
\begin{align}
  \nabla_a \Psi = n \nabla_a \mu_\tg + s \nabla_a T_\tg \ ,
\end{align}
we can add the two momentum equations to get the Euler equation 
\begin{align}
      \perp_a^b \nabla_b \Psi + \left(\Psi - \Lambda\right) u^b \nabla_b u_a & = 0 \ . \label{TOV}
\end{align}
We also anticipate the conservation of the specific entropy $x_\s = s/n$ along the worldlines of $u^a$. It follows from 
$\nabla_a n^a = \nabla_a s^a = 0$ that
\begin{align}
       u^a \nabla_a x_\s & = 0 \ . \label{xscons}
\end{align}

Now, 
eqs.~\eqref{TOV} and $\nabla_a n^a = \nabla_a s^a = 0$ are (almost!) the standard results for a non-dissipative, single-fluid, but 
double-constituent system. The main difference from the standard results is that we also have the remaining equation
\begin{align}
  \perp_a^b \nabla_b T_\tg + T_\tg u^b\nabla_b u_a = 0 \ , 
  \label{tolman}
\end{align} 
which follows from Eq.~\eqref{entcons}. This ``extra'' relation represents the system's two-fluid origin. Even though we have (by hand) reduced 
the four-velocities from two to one, the system ``remembers'' that it started out with two flux degrees of freedom, thus  leaving behind an extra 
field equation involving the entrainment $\Asn$. The additional equation represents a constraint on the evolution of the system. Later, in the specific examples 
considered in Sec.~\ref{spevar}, we will see that a (modified) constraint exists also when dissipation is incorporated. 

Before we proceed, it is important to understand that the presence of the additional constraint \eqref{tolman} should not come as a surprise. Consider, for example, (see Appendix~\ref{app:Tolman} for details) the case of a static and spherically symmetric spacetime with metric of form
\begin{align}
       d s^2 & = - e^{\nu\left(r\right)} d t^2 + e^{\lambda\left(r\right)} d r^2 + r^2 \left(d \theta^2 + \sin^2 \theta d \phi^2\right) \ ,
\end{align} 
and four-velocity given by
\begin{align}
      u^a & = \left(e^{- \nu / 2},0,0,0\right)  \ .
\end{align} 
In this case it is easy to show that \eqref{tolman} leads to \begin{subequations}
\begin{align}
      \frac{d}{d r} \left(\mu_\tg e^{\nu/2}\right) & = 0 \quad \Longrightarrow \quad \mu_\tg\left(r\right) = \mu_\tg\left(r_0\right) 
                           e^{- \left[\nu\left(r\right) - \nu\left(r_0\right)\right]/2} \ , \\
      \frac{d}{d r} \left(T_\tg e^{\nu/2}\right) & = 0 \quad \Longrightarrow \quad T_\tg\left(r\right) = T_\tg\left(r_0\right) 
                           e^{- \left[\nu\left(r\right) - \nu\left(r_0\right)\right]/2} \ , \label{tolrs}
\end{align} 
\end{subequations}
which are the well-known Tolman red-shift formulas \cite{tolman34:_book,landau1980:_statistical} for chemical potential and temperature. 
The first relation says that an observer at point $r$ will find---say, by inserting or taking away particles---that the chemical potential is
$\mu_\tg\left(r_0\right)$. However, an observer using the same process in the frame at $r$ will find the chemical potential to be  
$\mu_\tg\left(r_0\right) e^{- \left[\nu\left(r\right) - \nu\left(r_0\right)\right]/2}$. The main difference, compared to the classic result, is that \eqref{tolman} also applies to systems with nontrivial dynamics.

\subsection{``Generalized'' thermodynamics}
\label{gen} 

\noindent
Before we move on, it is worth explaining why we felt obliged to use the word ``generalized'' in the description of $\mu_a$, $\Theta_a$, and $\Psi$. The answer relates to the first law of thermodynamics for the system and tracks back to the existence of the two frames of reference. In a simple 
experiment, there is only the lab frame in which the total system under analysis is at rest. In that case we can define the set of (extensive) state 
parameters consisting of a total amount of heat (or entropy $S$), total volume $V$, and total number of particles $N$. We also can discover a 
total internal energy $U$ which is a function of all the state parameters; i.e.,~$U = U\left(S,V,N\right)$. 

The first law of thermodynamics states that an infinitesimal change in the total internal energy $d U$ can be written using a second set of 
(intensive) variables, which are the pressure $p$, the temperature $T$, and the chemical potential $\mu$. Using an Euler 
relation---$p + \varepsilon = \mu n + Ts$, $\varepsilon = U/V = \varepsilon\left(n,s\right)$, $n = N/V$, and $s = S/V$, \cite{reichl98:_book}---which exists because of 
scaling properties for this system, the first law can be written in terms of the densities, and takes the local form
\begin{align}
    d \varepsilon & = \frac{\partial \varepsilon}{\partial n} d n + \frac{\partial \varepsilon}{\partial s} d s \equiv \mu d n + T d s \quad \Longrightarrow \quad
    \mu = \frac{\partial \varepsilon}{\partial n}  \ , \quad T = \frac{\partial \varepsilon}{\partial s} \ .
    \label{trad1stlaw}
\end{align}

The two-fluid system is different in that the closest we have to a first law involves the fluxes:
\begin{align}
   \delta \Lambda & = \frac{\partial \Lambda}{\partial n^a} \delta n^a + \frac{\partial \Lambda}{\partial s^a} \delta s^a 
                                = \mu_a \delta n^a + \Theta_a \delta s^a  
                                \quad \Longrightarrow \quad
    \mu_a = \frac{\partial \Lambda}{\partial n^a} \ , \quad \Theta_a = \frac{\partial \Lambda}{\partial s^a} \ .
\end{align}
It is easy to see that, in this case  it is not possible to write down a definition for a  scalar chemical potential  that would agree with 
the traditional definition from Eq.~\eqref{trad1stlaw}. Similarly for the temperature. Because of this, $\Psi$ will not satisfy the standard Euler 
relation  but rather that given in Eq.~\eqref{genpress}, making $\Psi$ ``generalized'', as well. 
We can, of course, identify a term in $\mu_g = - u^a \mu_a$ that agrees with $\mu$ above, and likewise for $T_g = - \bu^a \Theta_a$:
\begin{subequations}
\begin{align}
    \Bn n & = \frac{\partial \rho}{\partial n} = \mu \ , \quad \Longrightarrow \quad \mu_g = \mu + \Asn s \ , \\
    \Bs s & = \frac{\partial \rho}{\partial s} = T  \quad \Longrightarrow \quad T_g = T + \Asn n \ .
\end{align}  
\end{subequations}
However, Eqs.~\eqref{singflmug} and \eqref{singflTg} then show that $\mu_g \neq \mu$ and $T_g \neq T$ even in the single-fluid limit.

\section{ Incomplete  Action Principle and Matter Space Fix}
\label{flact}

\noindent
The relativistic multifluid equations of motion can be derived from an action principle which uses the Lagrangian $\Lambda$. For the two-fluid system considered 
here, $\Lambda$ is a function of all the independent scalars which can be built using the spacetime metric $g_{a b}$, the particle flux $n^a$, 
and the entropy flux $s^a$. That is, the three independent scalars $n^2$, $s^2$ and $x^2$. The action is 
\begin{equation}
    S_F = \fint \Lambda\left(n^2,s^2,x^2\right) \ .
\end{equation}
A variation of $S_F$ with respect to $n^a$, $s^a$, and $g_{a b}$ leads to
\begin{align}
   \delta S_F & = \int_{\cal M} d^4 x \delta \left(\sqrt{- g} \Lambda\right) \cr
                     & = \int_{\cal M} d^4 x \sqrt{- g} \left[\mu_a \delta n^a + \Theta_a \delta s^a 
                      + \frac{1}{2} \left(\Lambda g^{a b} + \mu^a n^b + \Theta^a s^b\right) \delta g_{a b}\right] \ , \label{actvar}
\end{align}
where we have used the well-known fact that
\begin{align}
    \delta \sqrt{- g} & = \frac{1}{2} \sqrt{- g} g^{a b} \delta g_{a b} \ ,
\end{align}
and the chemical potential momentum $\mu_a$ and thermal momentum $\Theta_a$  have already been defined in 
Eqs.~\eqref{genmom} and \eqref{genTmom}, respectively.

At this point it is clear that the variation is incomplete because it implies for the general variations $\delta n^a$, $\delta s^a$, and $\delta g_{a b}$ that the 
equations of motion should be $\mu_a = 0 = \Theta_a$ and the $T^{a b}$ inferred from the coefficient of $\delta g_{a b}$ does not recover even the 
standard perfect fluid model \cite{mtw73}. 

\subsection{Matter space fix}

\noindent
In order to make the action principle viable, we need to introduce some additional infrastructure.
 The first step is to consider two abstract three-dimensional Riemannian manifolds---collectively called 
``matter spaces'' even though one is for the particles and the other relates to the entropy. \cite{Andersson15:_dissfl_act}). 
The two abstract spaces are taken to be independent of each other, which is consistent with the fact that the two flux degrees of freedom 
need to be varied independently in the action principle.

In the following, the points in the particle space manifold are labeled by coordinates $\xa$ ($A = 1, \ 2, \ 3$), and the entropy space points are labeled by  coordinates $\bxa$  ($\Ab = 1, \ 2, \ 3$). Both sets, $\xa$ and $\bxa$, represent scalar functions on spacetime. When repeated, say, with an index $A$ up 
and an index $A$ down, the matter space indices satisfy the Einstein summation convention. With this construction, each individual worldline of the field $u^a$ is 
mapped to an individual point $\xa$ in the particle space and likewise for the worldlines of $\bu^a$ and the points $\bxa$ in the entropy space. 
In spacetime the scalars $\xa$ and $\bxa$ vary from worldline to worldline but remain constant along the worldline to which they 
are initially assigned in the mapping process. In fact, any object that we write later, which carries only matter space indices, will be a scalar 
when evaluated on spacetime.

In the next section we will see how spacetime index carrying objects, like the three forms dual to $n^a$ and $s^a$, denoted $n_{a b c}$ and 
$s_{a b c}$, respectively, can be identified with a particle space index carrying three form $n_{A B C}$ and an entropy space three form 
$s_{\Ab \Bb \Cb}$, through use of the maps
\begin{align}
      \paa & = \nabla_a \xa \ , \quad \bpaa = \nabla_a \bxa \ .
\end{align}  
These maps can also be used to identify a particle space chemical potential  conjugate, $\mu^{A B C}$, to the particle three-form and an entropy space temperature conjugate, $\Theta^{\Ab \Bb \Cb}$. Inverting the map, this leads to spacetime quantities $\mu^{a b c}$ and  $\Theta^{a b c}$. The maps are essential components of nearly all the matter space degrees of 
freedom which can enter the action principle. 

\subsection{The Fundamental Variables}
\label{matsp_constrct}

\noindent
As just stated, we will use the $\paa$ and $\bpaa$ maps to ``pull-back/push-forward'' between index carrying objects in spacetime and index 
carrying objects in the matter spaces. To begin, we introduce the dual three forms $n_{a b c}$ and $s_{a b c}$:
\begin{subequations}\label{ns3frm}
\begin{align}
    n_{a b c} & = \epsilon_{d a b c} n^d \ , \quad n^a = \frac{1}{3!} \epsilon^{b c d a} n_{b c d} \ , \\
    s_{a b c} & = \epsilon_{d a b c} s^d \ , \quad s^a = \frac{1}{3!} \epsilon^{b c d a} s_{b c d} \ .
\end{align}
\end{subequations}
Next, we use the maps to relate the particle space three-form $n_{A B C}$ with $n_{a b c}$ and the entropy space three-form 
$s_{\Ab \Bb \Cb}$ with $s_{a b c}$; i.e.,,
\begin{subequations}
\begin{align}
    n_{a b c} & = \paa \pbb \pcc n_{A B C} \ , \label{part3form} \\
    s_{a b c} & = \bpaa \bpbb \bpcc s_{\Ab \Bb \Cb} \ . \label{ent3form}
\end{align}
\end{subequations}
Similarly, we introduce the dual objects for $\mu_a$ and $\Theta_a$; i.e.,
\begin{subequations}
\begin{align}
    \mu^{a b c} & = \epsilon^{d a b c} \mu_d \ , \quad \mu_a = \frac{1}{3!} \epsilon_{b c d a} \mu^{b c d} \ , \\
    \Theta^{a b c} & = \epsilon^{d a b c} \Theta_d \ , \quad \Theta_a = \frac{1}{3!} \epsilon_{b c d a} \Theta^{b c d} \ .
\end{align}
\end{subequations}
These explicit examples show how the double-index structure from two different spaces (one or the other of $\{A,\Ab\}$ for the matter 
space and $a$ for spacetime) leads to the operations of pull-back/push-forward induced by the maps $\paa$ and $\bpaa$.

For the particle space there is an antisymmetric volume form $\epsilon_{B C D}$ and for the entropy space there is an analogous  
$\bar{\epsilon}_{\Bb \Cb \Db}$ (each having their own Jacobian factors; see \cite{andersson24:_ent1diss} for details). These allow us to write 
$n_{A B C} = n \epsilon_{B C D}$ and $s_{\Ab \Bb \Cb} = s \bar{\epsilon}_{\Bb \Cb \Db}$. Using Eq.~\eqref{ns3frm} to rewrite $n^a$ and 
$s^a$ in terms of their respective three forms $n_{a b c}$ and $s_{a b c}$, and then substituting into these rewritten terms the relations in 
Eqs.~\eqref{part3form} and \eqref{ent3form}, we can show that $u^a = n^a/n$ and $\bu^a = s^a/s$ leads to
\begin{equation}
    u^a = \frac{1}{3!} \epsilon^{b c d a} \pbb \pcc \pdd \epsilon_{B C D} \ , \quad \bu^a = \frac{1}{3!} \epsilon^{b c d a} 
              \bpbb \bpcc \bpdd \bar{\epsilon}_{\Bb \Cb \Db} \ ,
\end{equation}
Because the two fluids are not parallel, a given $u^a$ worldline will be intersected 
by a continuum of worldlines of $\bu^a$, and vice versa. The idea is illustrated in Figure~2 of \cite{Andersson15:_dissfl_act}.

Since there are two independent unit four-velocities, there are two natural notions of ``time-derivative'': ${\cal L}_u$ and ${\cal L}_{\bu}$, where
${\cal L}_\zeta$ is the Lie derivative with respect to a vector field $\zeta^a$. We note the important consistency checks that
\begin{equation}
     u^a \paa = u^a \nabla_a \xa = {\cal L}_u \xa = 0 \ , \quad \bu^a \bpaa = {\cal L}_{\bu} \bxa = 0 \ ,
     \label{xliedrag}
\end{equation} 
which must hold because the map $\paa$ is contracted four times on $\epsilon^{b c d a}$ but $\xa$ has only three components; similarly for 
$\bu^a \bpaa$. This shows that the map $\paa$ is everywhere orthogonal to the worldlines of $u^a$ and likewise for $\bpaa$ along the $\bu^a$. 
This leads to the immediate demonstration that the $\Psi^A_0$ and $\Psi^A_i$ columns are not linearly independent (and the same for 
$\bpaa$):
\begin{subequations}
\begin{align}
    \paa u^a & = 0 \quad \Longrightarrow \quad \Psi^A_0 = - \Psi^A_i \frac{u^i}{u^0} \ , \\
    \bpaa \bu^a & = 0 \quad \Longrightarrow \quad \bar{\Psi}^A_0 = - \bar{\Psi}^A_i \frac{\bu^i}{\bu^0} \ .
\end{align}
\end{subequations}

Another way to describe the conditions in Eq.~\eqref{xliedrag} is to say that the matter space coordinates are Lie-dragged along their 
respective fluid worldlines. This might have been expected as the basic role of the maps is to identify all the spacetime points on a specific 
worldline with a specific point in the relevant matter space. Because $u^a$ and $\bu^a$ are not parallel, their worldlines cut across each other, 
meaning generally that the $\xa$ are not Lie-dragged along $\bu^a$ and the $\bxa$ are not Lie-dragged along the $u^a$. This leads to a new set of matter space 
``vectors'':  
\begin{align}
   V^A & = \bu^c \pac = {\cal L}_{\bu} \xa \ , \quad \bar{V}^\Ab = u^c \bpac = {\cal L}_u \bxa \ .
\end{align}
Unlike the previous set $\{n_{A B C} ,s_{\Ab \Bb \Cb}\}$, these matter space degrees of freedom vanish when the two fluids are comoving, 
i.e., when $u^a = \bu^a$. They did not show up in the two previous discussions \cite{Andersson15:_dissfl_act,andersson24:_ent1diss}, because 
\cite{Andersson15:_dissfl_act} did not consider derivatives and \cite{andersson24:_ent1diss} considered only a single fluid, and so both quantities would automatically vanish. Even though we refer to these as ``velocities'' in the following; strictly speaking,  $V^A$ represents the change of $\xa$ along $\bu^a$ while 
$\bar{V}^\Ab$ represents the change in $\bxa$ along $u^a$.

The remaining dynamical spacetime  field is the metric $g^{a b}$. Using the maps $\paa$ and $\bpaa$ we may construct three matter space 
``metrics'' $\gab$, $\bgab$, and $\hgab$:
\begin{subequations}
\begin{align}
      \gab & = \paa \pbb g^{a b} = \gba \ , \\
      \bgab & = \bpaa \bpbb g^{a b} = \bgba \ , \\ 
      \hgab & = \paa \bpbb g^{a b} = \bpba \pab g^{a b} = \hgba \ .
\end{align}
\end{subequations}
Their inverses $\ginvab$, $\bginvab$, and $\hginvab$ can be defined in the usual Linear Algebra way (Cramer's Rule \cite{strang80:_lin_alg}), 
and explicit formulations can be found in \cite{andersson24:_ent1diss}. These represent the third set of matter space degrees of 
freedom. Like the terms presented before, all three ``metrics'' are scalars when evaluated on spacetime.

Before moving on, we should say a few words about the meaning and use of the set of indices $\{A,\Ab\}$. We started off by 
assuming we have two $3D$ Riemannian spaces, one with the coordinates $\xa$ and the other with the coordinates $\bxa$. 
We  also introduced two sets of 
functions: $\left(n_{A B C},\mu^{A B C}\right)$ for the particle space and $\left(s_{\Ab \Bb \Cb},\Theta^{\Ab \Bb \Cb}\right)$ for the entropy 
space. When there is no dissipation, the only type of sums we need for the matter space indices are ones which perform an Einstein sum over 
objects which have only indices from the particle space, such as $n_{A C D} \mu^{B C D}$, or only indices from the entropy space, such as 
$s_{\Ab \Bb \Eb} \Theta^{\Cb \Db \Eb}$. 
However, when we consider dissipation, it is necessary to break the separation, by allowing particle space indices to have an 
Einstein summation convention with entropy space indices. As an example, we might have $n_{A B C} \Theta^{\Db \Eb \Cb}$, where the $C$ and $\Cb$ are summed. 

The meaning of the required kind of summation can be understood by a simple example. 
 Assume that $V^a$ is a vector that is spatial with respect to \emph{both} maps. We then have
\begin{equation}
    V^c V_c = g^{bc} V_b V_c 
    = g^{bc} \Psi^B_b V_B \bar{\Psi}^{\bar{C}}_c V_{\bar{C}} 
    = \hat{g}^{B\bar{C}} V_B V_{\bar{C}} \, ,
\end{equation}
and we simply \emph{call} this $V^C V_{\bar{C}}$. For a general vector its meaning is ``the norm of the part of the vector $V^a$ that is spatial as seen by \emph{both} maps''. The idea extends in the usual way to higher order tensors.

Another important point raised in \cite{andersson24:_ent1diss} is that time derivatives of the matter space metrics must be incorporated in the 
action principle; otherwise, the anticipated results for shear and bulk viscosity cannot be incorporated (see the discussion in \cite{celora21:_lindiss}). Because of the two notions of time 
derivatives, we need to consider six metric derivatives: $\{\dgab,\dbugab\}$, $\{\bdgab,\dbubgab\}$, and $\{\hdgab,\dhubgab\}$. From the 
definitions of the matter space metrics we find 
\begin{subequations}
\begin{align}
   \dgab & = - 2 \paa \pbb \nabla^{(a} u^{b)} \ , \label{dgab1} \\ 
   \dbubgab & = - 2 \bpaa \bpbb \nabla^{(a} \bu^{b)} \ , \label{dgab4} \\
   \bdgab & = \bpbb \nabla^b \left(u^a \bpaa\right) + \bpaa \nabla^a \left(u^b \bpbb\right) - 2 \bpaa \bpbb \nabla^{(a} u^{b)} \ , \label{dgab3} \\
   \dbugab & = \pbb \nabla^b \left(\bu^a \paa\right) + \paa \nabla^a \left(\bu^b \pbb\right) - 2 \paa \pbb \nabla^{(a} \bu^{b)} \ ,  \label{dgab2} \\
   \hdgab & =  \paa \nabla^a \left(u^b \bpbb\right) - 2 \paa \bpbb \nabla^{(a} u^{b)}  \ , \label{dgab5} \\
   \dhubgab & = \bpbb \nabla^b \left(\bu^a \paa\right) - 2 \paa \bpbb \nabla^{(a} \bu^{b)} \ . \label{dgab6}                   
\end{align}
\end{subequations}
Here we have made use of the fact that $\nabla_a \nabla_b \xa = \nabla_b \nabla_a \xa$, and similarly for derivatives of $\bxa$. These matter 
space metric derivatives complete the set of matter space variables to be included in our analysis. Like their predecessors, they are scalars when evaluated on spacetime. 

In summary, we have shown that the maps $\paa$ and $\bpaa$ acting on the spacetime degrees of freedom $\{n^a,s^a,g_{a b}\}$, and 
taking Lie derivatives of the matter space metrics, leads to the following sets of fundamental, matter space variables:\footnote{We acknowledge the existence of ``acceleration'' terms $\{{\cal L}_u \bar{V}^\Ab,{\cal L}_u V^A,{\cal L}_{\bu} \bar{V}^\Ab,{\cal L}_{\bu} V^A\}$, but have chosen to leave a study of their impact for later work, so as to limit the focus here on elements that are most likely to contribute to the dissipation channels of resistivity, and bulk and shear viscosities.}  
\begin{subequations}
\begin{align}
      {\cal S}_X & = \left\{\bxa,\xa\right\} \ , \label{SX} \\ 
      {\cal S}_V & = \left\{\bar{V}^\Ab,V^A\right\} \ , \label{SV} \\
      {\cal S}_g & = \{\bgab,\gab,\hgab\} \ , \label{Sg} \\
      {\cal S}_{Lg} & = \left\{\dbubgab,\dbungab,\dbugab,\dgab,\dhubgab,\hdgab\right\} \ . \label{SLg}
\end{align} 
\end{subequations}
Locally (on matter space) these objects transform as tensors. 

\subsection{Connecting Spacetime Fibrations with \texorpdfstring{${\cal S}_g$}{Matter Space Metrics} and 
\texorpdfstring{${\cal S}_{Lg}$}{Matter Space Metric Derivatives}}
\label{uafib}

\noindent
It is well-established in the literature on general relativistic fluids that the form $\nabla^a u^b$ is the principal object that determines different channels of dissipation (bulk, shear, etc.). The various channels are extracted through the use of the standard
decomposition of $\nabla^a u^b$; namely, for the particle flow we have
\begin{subequations}
\begin{align}
   \nabla^a u^b & = \sigma^{a b} + \frac{1}{3} \theta \perp^{a b} + \varpi^{a b} - u^a a^b \ , \\
   \sigma^{a b} & = \frac{1}{2} \left(\perp^{a c} \nabla_c u^b + \perp^{b c} \nabla_c u^a\right) - \frac{1}{3} \theta \perp^{a b} 
        = \sigma^{b a} \ , \\
   \varpi^{a b} & = \frac{1}{2} \left(\perp^{a c} \nabla_c u^b - \perp^{b c} \nabla_c u^a\right) = - \varpi^{b a} \ , \\
   a^a & = u^b \nabla_b u^a \ , \\
   \theta & = \nabla_a u^a \ . 
\label{nabu}
\end{align}
\end{subequations}
Similarly, for the entropy flow we have
\begin{subequations}
\begin{align}
   \nabla^a \bu^b & = \bar{\sigma}^{a b} + \frac{1}{3} \bar{\theta} \bar{\perp}^{a b} + \bar{\varpi}^{a b} - \bu^a  
                                       \bar{a}^b \ , \\ 
   \bar{\sigma}^{a b} & = \frac{1}{2} \left(\bar{\perp}^{a c} \nabla_c \bu^b + \bar{\perp}^{b c} \nabla_c \bu^a\right) 
                                       - \frac{1}{3} \bar{\theta} \bar{\perp}^{a b} = \bar{\sigma}^{b a} \ , \\ 
   \bar{\varpi}^{a b} & = \frac{1}{2} \left(\bar{\perp}^{a c} \nabla_c \bu^b - \bar{\perp}^{b c} \nabla_c \bu^a\right) 
                                  = - \bar{\varpi}^{b a} \ , \\ 
   \bar{\perp}^{a b} & = g^{a b} + \bu^a \bu^b = \bar{\perp}^{b a} \ , \\ 
   \bar{a}^a & = \bu^b \nabla_b \bu^a \ , \\
   \bar{\theta} & = \nabla_a \bu^a \ . 
\end{align}
\end{subequations}
Obviously $\perp^{a b} u_b = \perp^{b a} u_b = 0$, which means $\sigma^{a b} u_b = \sigma^{b a} u_b = 0$ and 
$\varpi^{a b} u_b = - \varpi^{b a} u_b = 0$, as well. It is also the case that $\perp^{a b} \sigma_{a b} = 0$ and $\perp^{a b} \varpi_{a b} = 0$. Finally, 
because $u_a u^a = - 1$, we have $u^a a_a = 0$. Analogous conclusions hold for the entropy fields.

Here, the $\perp_{a b}$ projection is directly connected with $\ginvab$ and $\bar{\perp}_{a b}$ with $\bginvab$ (see \cite{andersson24:_ent1diss} for details) 
since
\begin{align}
    \paa \pbb \ginvab & = \perp_{a b} \ , \quad \bpaa \bpbb \bginvab = \bar{\perp}_{a b} \ .
\end{align}
It is straightforward to see, via simple substitution, that
\begin{subequations}
 \begin{align}
   \dgab & = - 2 \paa \pbb \left(\sigma^{a b} + \frac{1}{3} \theta \perp^{a b}\right) \ , \label{2dgab1} \\ 
   \dbubgab & = - 2 \bpaa \bpbb \left(\bar{\sigma}^{a b} + \frac{1}{3} \bar{\theta} \bar{\perp}^{a b}\right) \ . \label{2dgab4}                    
\end{align}
\end{subequations}
If we contract both sides of Eq.~\eqref{dgab1} with $\ginvab$, and Eq.~\eqref{dgab4} with $\bginvab$, we have, respectively,
\begin{subequations}
\begin{align}
      \ginvab \dgab & = - 2 \theta \ , \label{theg} \\
      \bginvab \bdgab & = - 2 \bar{\theta} \ . \label{btheg}
\end{align}
\end{subequations}
These relations are important as they indicate how matter space metric time derivatives in the action principle can lead to  shear and bulk 
viscous dissipation, since after the equations of motion are derived one can, say, substitute in for $\dgab$ the right-hand-side of 
Eq.~\eqref{2dgab1}.

\section{ Lagrangian Displacements}
\label{lagdis}

\noindent
At this point we can gather together the main elements required to build the changes $\delta n^a$, $\delta s^a$, and $\delta g_{a b}$ so that we 
have a viable variation of the fluid action given in Eq.~\eqref{actvar}. In order to arrive at the correct equations of motion we need to introduce 
the Lagrangian variations $\Delta = \delta + {\cal L}_\xi$ and $\bar{\Delta} = \delta + {\cal L}_{\bar{\xi}}$, where $\delta$ is an Eulerian variation 
and ${\cal L}_\xi$ and ${\cal L}_{\bar{\xi}}$ are the Lie derivatives along, respectively, the spacetime displacements $\xi^a$ and $\bar{\xi}^a$. 

The most tedious part of the calculation involves building the Lagrangian variations of matter space degrees of freedom and then using them 
in the action principle. We have worked out the full suite of changes of the matter space variables $\{{\cal S}_X,{\cal S}_V,{\cal S}_g\}$, and 
will provide those results in the following. We have, however, worked out only the $\bar{\Delta}$ variations for the set ${\cal S}_{Lg}$. The reason for completely working 
out the first set is that their $\Delta$ and $\bar{\Delta}$ variations are required to derive the more 
expansive variations of ${\cal S}_{Lg}$. The reason we need only the $\bar{\Delta}$ variations of ${\cal S}_{Lg}$ will be stated below. Moreover, we will only explicitly provide 
$\bar{\Delta} \dgab$  in this section with the remaining quantities relegated to Appendix~\ref{appen}. 

\subsection{Lagrangian Variations of the Matter Space Coordinates \texorpdfstring{${\cal S}_X$}{Matter Space Coordinates}, the 
Velocities \texorpdfstring{${\cal S}_V$}{Matter Space Velocities}, and Metrics \texorpdfstring{${\cal S}_g$}{Matter Space Metrics}}

\noindent
Let us first establish a more intuitive understanding of the Lagrangian displacements $\xi^a$ and $\bar{\xi}^a$. Pick an unvaried worldline 
$u^a_I$ and at some initial point on it attach the tail of a Lagrangian variation $\xi^a_i$, $i = \ 1 \ , 2 \dots N$; proceed to the next point on 
the $u^a_I$ worldline and do the same thing; continue until you reach the point at $i = N$. If we suppose that this process has been done 
smoothly so as to prevent discontinuous jumps in the tips, then the curve made by drawing a line from the first tip to the next, and so on, will 
form a smooth curve that represents the varied worldline of the original $u^a_I$. The operator $\delta$ is Eulerian since it prescribes changes in fields at 
fixed spacetime points, whereas the Lagrangian operators $\Delta$ and $\bar{\Delta}$ measure differences by comparing a field's value at the 
tail of a Lagrangian displacement with its value at the tip.

When a worldline is varied it must still be the case that its own $\xa$ (or $\bxa$) remains fixed; that is, observers holding on to different fixed 
points on the worldline will not see $\xa$ (or $\bxa$) change as they are taken along their own tail-to-tip displacement. The implication, then, is 
that $\delta \xa$ and $\delta \bxa$ and $\xi^a$ and 
$\bar{\xi}^a$ must be such that they lead to $\Delta \xa = 0$ and $\bar{\Delta} \bxa = 0$. Hence, we find  
\begin{subequations}
\begin{align}
    \delta \xa & = - {\cal L}_\xi \xa = - \xi^a \partial_a \xa = - \paa \xi^a \ , \\
    \delta \bxa & = - {\cal L}_{\bar \xi} \bxa = - \bar{\xi}^a \partial_a \bxa = - \bpaa \bar{\xi}^a \ .    
\end{align}
\end{subequations}
When we consider the action principle, we are then looking for the conditions (field equations) that lead to $\delta S_F = 0$ for variations 
$\delta \xa$ (actually $\xi^a$) and $\delta \bxa$ (actually $\bar{\xi}^a$), which are subject only to the constraints $\Delta \xa = 0$ and 
$\bar{\Delta} \bxa = 0$. We will be able to use these constraints to ``fix'' the variations $\delta n^a$ and $\delta s^a$ so that the action principle delivers 
viable fluid equations of motion and a $T^{a b}$ tensor that can be inserted into the Einstein equations to determine the gravitational field.

We need to determine $\bar{\Delta} \xa$ and $\Delta \bxa$. When $u^a$ and $\bu^a$ are not parallel, and assuming no phase 
separation, then in principle every $u^a$ worldline is intersected at each of its points by an individual worldline from the field $\bu^a$, and vice 
versa for $\bu^a$ and the field $u^a$. The operation $\Delta \bxa$ takes the value $\bxa$ of the worldline which intersects the non-displaced 
worldline of $u^a$ at the tail of $\xi^a$ and then carries it along to the tip of $\xi^a$ on the displaced $u^a$ worldline. Using $\Delta \xa = 0$ 
and $\bar{\Delta} \bxa = 0$, we can write 
\begin{subequations}
\begin{align}
       \bar{\Delta} \xa & = \delta \xa + \bar{\xi}^a \nabla_a \xa = - \paa \xi^a + \paa \bar{\xi}^a = \paa \left(\bar{\xi}^a - \xi^a\right) \ , \\
       \Delta \bxa & = \delta \bxa + \xi^a \nabla_a \bxa = - \bpaa \bar{\xi}^a + \bpaa \xi^a = \bpaa \left(\xi^a - \bar{\xi}^a\right) \ .
\end{align}
\end{subequations}
To determine the Lagrangian variations of $\paa$ and $\bpaa$ we note that $\delta \nabla_a \xa = \nabla_a \delta \xa$ (similarly for $\bxa$) and therefore find 
\begin{subequations}
\begin{align}
        \Delta \paa & = 0 \ , \quad
        \bar{\Delta} \paa = \nabla_a \left[\pab \left(\bar{\xi}^b - \xi^b\right)\right] \ , \\
        \bar{\Delta} \bpaa & = 0 \ , \quad
        \Delta \bpaa = \nabla_a \left[\bpab \left(\xi^b - \bar{\xi}^b\right)\right] \ .
\end{align}
\end{subequations}
The meaning of $\bar{\Delta} \paa$ and $\Delta \bpaa$ is completely analogous to the description of $\bar{\Delta} \xa$ and $\Delta \bxa$ 
above. This understanding is carried throughout this section.

Using the above Lagrangian variations of $\paa$ and $\bpaa$, we find that applying $\Delta$ and $\bar{\Delta}$ to $V^A$ and $\bar{V}^{\Ab}$ leads to
\begin{subequations}
\begin{align}
    \Delta V^A & = \paa \left[\bu^b \nabla_b \left(\bar{\xi}^a - \xi^a\right) - \frac{1}{2} \bu^a \bu_b \bu_c \bar{\Delta} g^{b c} 
                                     - \left(\bar{\xi}^b - \xi^b\right) \nabla_b \bu^a\right] \ , \\
    \bar{\Delta} V^A & = \paa \left[\bu^b \nabla_b \left(\bar{\xi}^a - \xi^a\right) - \frac{1}{2} \bu^a \bu_b \bu_c \bar{\Delta} g^{b c}\right] 
                                     + \left(\nabla_b \paa\right) \bu^a \left(\bar{\xi}^b - \xi^b\right) \ , \\
    \Delta \bar{V}^\Ab & = \bpaa \left[u^b \nabla_b \left(\xi^a - \bar{\xi}^a\right) - \frac{1}{2} u^a u_b u_c \Delta g^{b c}\right]  
                                     + \left(\nabla_b \bpaa\right) u^a \left(\xi^b - \bar{\xi}^b\right) \ , \\
    \bar{\Delta} \bar{V}^\Ab & = \bpaa \left[u^b \nabla_b \left(\xi^a - \bar{\xi}^a\right)  - \frac{1}{2} u^a u_b u_c \Delta g^{b c} 
                                      - \left(\xi^b - \bar{\xi}^b\right) \nabla_b u^a\right] \ . 
\end{align}
\end{subequations}
Applying the same variations to the definitions for $\gab$, $\bgab$, and $\hgab$, we have
\begin{subequations}
\begin{align}
    \Delta \gab & = \paa \pbb \Delta g^{a b} \ , \\
    \bar{\Delta} \gab & = \paa \pbb \Delta g^{a b} + \left(\paa \nabla_c \pbb + \pbb \nabla_c \paa\right) g^{a b} \left(\bar{\xi}^c - \xi^c\right) \ , \\
    \Delta \bgab & = \bpaa \bpbb \bar{\Delta} g^{a b} + \left(\bpaa \nabla_c \bpbb + \bpbb \nabla_c \bpaa\right) g^{a b} 
                                \left(\xi^c - \bar{\xi}^c\right) \ , \\
    \bar{\Delta} \bgab & = \bpaa \bpbb \bar{\Delta} g^{a b} \ , \\
    \Delta \hgab & = \paa \bpbb \left[\Delta g^{a b} + \nabla^a \left(\xi^b - \bar{\xi}^b\right)\right] + \left(\paa \nabla_c \bpbb\right) g^{a b} 
                                 \left(\xi^c - \bar{\xi}^c\right) \ , \\
    \bar{\Delta} \hgab & =  \paa \bpbb \left[\bar{\Delta} g^{a b} + \nabla^b \left(\bar{\xi}^a - \xi^a\right)\right] + \left(\bpbb \nabla_c \paa\right) 
                                 g^{a b} \left(\bar{\xi}^c - \xi^c\right) \ .       
\end{align}
\end{subequations}
Here, we have used the following variations of the four-velocities $u^a$ and $\bu^a$ (see \cite{andersson24:_ent1diss} for details):
\begin{subequations}
\begin{align}
     \Delta \bu^a & = - \frac{1}{2} \left(\bu_b \bu_c \bar{\Delta} g^{b c}\right) \bu^a + \left(\xi^b - \bar{\xi}^b\right) \nabla_b \bu^a 
                                + \bu^b \nabla_b \left(\bar{\xi}^a - \xi^a\right) \ , \\
     \bar{\Delta} \bu^a & = - \frac{1}{2} \left(\bu_b \bu_c \bar{\Delta} g^{b c}\right) \bu^a \ , \\
     \Delta u^a & = - \frac{1}{2} \left(u_b u_c \Delta g^{b c}\right) u^a \ , \\
     \bar{\Delta} u^a & = - \frac{1}{2} \left(u_b u_c \Delta g^{b c}\right) u^a + \left(\bar{\xi}^b - \xi^b\right) \nabla_b u^a 
                                + u^b \nabla_b \left(\xi^a - \bar{\xi}^a\right) \ . \label{Delua}
\end{align}
\end{subequations}
Later we will also make use of the essential relations 
\begin{subequations}
\begin{align}
       \Delta g^{a b} &= \delta g^{a b} - 2 \nabla^{(a} \xi^{b)} \\
      \bar{\Delta} g^{a b} &= \delta g^{a b} - 2 \nabla^{(a} \bar{\xi}^{b)} \ .
\end{align}
\end{subequations}

\subsection{Lagrangian Variations of the Matter Space Metric Derivatives \texorpdfstring{${\cal S}_{Lg}$}{Matter Space Metric Derivatives}}

\noindent
We have already discussed that, having two sets of worldlines means there are two distinct Lagrangian variations: $\Delta$ and $\bar{\Delta}$. 
When we consider the variations of fields in the action principle, we have to consider both variations for each and every 
field. Otherwise we do not retain all the dynamical degrees of freedom. However, it follows from the chain rule that when the action principle is imposed, a variation of the particle flux that begins with $\Delta n_{A BC}$, will 
have terms varied only by $\Delta$, and a variation of the entropy flux of the form $\bar{\Delta} s_{\Ab \Bb \Cb}$ will have terms varied only by 
$\bar{\Delta}$. 

While we have worked out both variations for the variables considered thus far---$\{{\cal S}_X,{\cal S}_V,{\cal S}_g\}$, we will only write down the $\bar{\Delta}$ variation of 
$\dgab$ in what follows. This is the only degree of freedom from ${\cal S}_{Lg}$ that we will consider in the discussion (for completeness, the other 
variations of ${\cal S}_{Lg}$ can be found in Appendix~\ref{appen}). 

The starting point for constructing the $\bar{\Delta}$ variation of $\dgab$ is
\begin{align}
   \bar{\Delta} \dgab & = \left(\nabla_a \gab\right) \bar{\Delta} u^a + u^a \bar{\Delta} \left(\nabla_a \gab\right) \ . 
\end{align}
Fortunately, $\bar{\Delta}$ and $\nabla_a$ commute when acting on $\bgab$, $\gab$, $\hgab$; that is, for $\gab$ we have
\begin{align}
     \bar{\Delta} \left(\nabla_a \gab\right) & = \nabla_a \left(\bar{\Delta} \gab\right) \ , \label{delpargab_maintext}
\end{align}
and likewise for the rest. Putting in the various pieces worked out above, we arrive at
\begin{align}        
  \bar{\Delta} \dgab & = \paa \pbb \left\{u^c \nabla_c \Delta g^{a b} + \left[u_c u_d \nabla^{(a} u^{b)} 
              - 2 \perp^{(a}_c \perp^{b)}_e \nabla_d u^e\right] \Delta g^{c d}\right\}  \cr
              & + \left(\paa \nabla_c \pbb\right) \left[g^{a b} \left(\nabla_d u^c\right) \left(\bar{\xi}^d - \xi^d\right) - \left(\nabla^b u^a\right) 
              \left(\bar{\xi}^c - \xi^c\right)\right] \cr
              & + \left(\pbb \nabla_c \paa\right) \left[g^{a b} \left(\nabla_d u^c\right) \left(\bar{\xi}^d - \xi^d\right) - \left(\nabla^a u^b\right) 
              \left(\bar{\xi}^c - \xi^c\right)\right] \cr
             & + \paa \nabla_c \left(\nabla_d \pbb\right) g^{a b} u^c \left(\bar{\xi}^d - \xi^d\right) 
             + \pbb \nabla_d \left(\nabla_c \paa\right) g^{a b} u^d \left(\bar{\xi}^c - \xi^c\right) \ . 
\label{dellie1}    
\end{align}

\section{The Spacetime Field Variations}
\label{fldeqs}

The key to including  dissipation in the variational formulation lies in the specific functional dependencies of $n_{ABC}$ and 
$s_{\Ab \Bb \Cb}$  \cite{Andersson15:_dissfl_act}. To illustrate the complexity this can lead to we will write down the most general form for $s_{\Ab \Bb \Cb}$ here, but restrict ourselves to  
fewer degrees of freedom later.  We are also simplifying the model by assuming a single, conserved, matter current. The specific forms for $n_{ABC}$ and $s_{\Ab \Bb \Cb}$ are then given by
\begin{subequations}
\begin{align}
    n_{ABC} & = n_{A B C}\left(\xd\right) \ , \label{nfuncdep} \\
    s_{\Ab \Bb \Cb} & =  s_{\Ab \Bb \Cb}\left({\cal S}_X,{\cal S}_V,{\cal S}_g,{\cal S}_{Lg}\right) \ ,
\label{sfuncdep}
\end{align}
\end{subequations}
where the variables required for $s_{\Ab \Bb \Cb}$ are defined in Eqs.~\eqref{SX} -- \eqref{SLg}. 

It should be  evident that $\Delta n_{A B C} = 0$, since it only depends on $\xa$; i.e.,, using the form given in Eq.~\eqref{nfuncdep}:
\begin{align}
   \Delta n_{A B C} & = \frac{\partial n_{A B C}}{\partial \xd} \Delta \xd = 0 \ . 
\end{align}
Consequently, the particle flux creation rate, $\Gamma_\text{n}$, vanishes since $\nabla_{[a} \Psi_{b]}^B = 0$, etc.: 
\begin{align}
   \Gamma_\mathrm{n} & =  \nabla_a n^a  = \frac{1}{3!} \epsilon^{b c d a} \nabla_{[a} n_{b c d]} \cr
     & = \frac{1}{3!} \epsilon^{b c d a} \nabla_{[a} \left(\Psi_b^B \Psi_c^C \Psi_{d]}^D n_{B C D}\right) \cr
     & = - \frac{1}{3!} \epsilon^{b c d a} \Psi_{[b}^B \Psi_c^C \Psi_d^D \Psi_{a]}^A \frac{\partial n_{B C D}}{\partial X^A} 
     \equiv 0 \ .
\end{align}
However, the extra dependencies for $s_{\Ab \Bb \Cb}$ breaks the closure of the entropy three-form $s_{a b c}$ in Eq.~\eqref{sfuncdep}, 
which leads to a non-zero entropy creation rate, $\Gamma_\text{s} = \nabla_a s^a \neq 0$, and hence the system is dissipative.

\subsection{Construction of \texorpdfstring{$\delta n^a$}{the Number Flux Variation}}

\noindent
To work out $\delta n^a$, we first note that since $\Delta n_{A B C} = 0$ and $\Delta \paa = 0$, we have $\Delta n_{a b c} = 0$ and therefore
\begin{align}
    \delta n_{a b c} & = - {\cal L}_\xi n_{a b c} \ .
\end{align}
Noting that
\begin{align}
    \frac{1}{3!} \epsilon^{b c d a} {\cal L}_\xi n_{b c d} & = \xi^b \nabla_b n^a - \left(n^d \nabla_d \xi^a 
    - n^a \nabla_d \xi^d\right) \ ,
\end{align} 
we see that
\begin{align}
    \delta n^a & = - \left[\xi^b \nabla_b n^a - \left(n^d \nabla_d \xi^a - n^a \nabla_d \xi^d\right)\right] 
                           + \frac{1}{2} n^a g_{b c} \delta g^{b c} \ , 
\end{align} 
where we have used
\begin{align}
    \delta \epsilon^{b c d a} & = \frac{1}{2} \epsilon^{b c d a} g_{e f} \delta g^{e f} \ .
\end{align}

For the action principle we have
\begin{align}
    \mu_a \delta n^a & = \left(- 2 n^b \nabla_{[b} \mu_{a]}\right) \xi^a - \frac{1}{2} \mu_c n^c g^{a b} \delta g_{a b} + {\cal BT} \ , 
    \label{nconserv}
\end{align} 
where ${\cal BT}$ represents terms that are total derivatives which neither contribute to the equations of motion nor to $T^{a b}$. Note that Eq.~\eqref{nconserv} allows one to read off the field equation (the coefficient 
of $\xi^a$) discussed earlier in Sec.~\ref{zerodissfluids}.

\subsection{Construction of \texorpdfstring{$\delta s^a$}{the Entropy Flux Variation}}

\noindent
To perform the corresponding calculation for $\delta s^a$ we need to be explicit about the functional dependence of $\bar{\Delta} s_{\Ab \Bb \Cb}$. We will be 
using the following subset of terms from $\left\{{\cal S}_X, {\cal S}_V, {\cal S}_g, {\cal S}_{Lg}\right\}$, that is,
$\{\xa,V^A,\bar{V}^\Ab,\gab,\bgab,\dgab\}$. These are the degrees of freedom which will be used later in Sec.~\ref{spevar} when specific 
actions are explored. For this subset of variables, the chain rule implies
\begin{align}
   \bar{\Delta} s_{\Ab \Bb \Cb} & = \frac{\partial s_{\Ab \Bb \Cb}}{\partial \xd} \bar{\Delta} \xd + \frac{\partial s_{\Ab \Bb \Cb}}{\partial V^D} 
   \bar{\Delta} V^D + \frac{\partial s_{\Ab \Bb \Cb}}{\partial \bar{V}^\Db} \bar{\Delta} \bar{V}^\Db + 
   \frac{\partial s_{\Ab \Bb \Cb}}{\partial \bar{g}^{\Db \Eb}} \bar{\Delta} \bar{g}^{\Db \Eb} 
   + \frac{\partial s_{\Ab \Bb \Cb}}{\partial \gde} \bar{\Delta} \gde \cr
   & + \frac{\partial s_{\Ab \Bb \Cb}}{\partial \dgde} \bar{\Delta} \dgde \ ,
   \label{delbsabc}
\end{align}
where we have used the fact that $\bar{\Delta} \bxd = 0$. Because there are six terms in Eq.~\eqref{delbsabc}, we will introduce the notation
\begin{align}
   \bar{\Delta} s_{\Ab \Bb \Cb} & = \sum_{I = 1}^6 \left.\bar{\Delta} s_{\Ab \Bb \Cb}\right|_I \ ,
\end{align}
where $\left.\bar{\Delta} s_{\Ab \Bb \Cb}\right|_1 \propto \bar{\Delta} \xd$ is the first term in Eq.~\eqref{delbsabc}, 
$\left.\bar{\Delta} s_{\Ab \Bb \Cb}\right|_2 \propto \bar{\Delta} V^A$ is the second, and so on to 
$\left.\bar{\Delta} s_{\Ab \Bb \Cb}\right|_6 \propto \bar{\Delta} \dgde$, which is the last term in Eq.~\eqref{delbsabc}.

Recalling that $\bar{\Delta} \bpaa = 0$, we see that
\begin{equation}
      \bar{\Delta} s_{a b c} = \bar{\Psi}_a^{[\Ab} \bar{\Psi}_b^\Bb \bar{\Psi}_c^{\Cb]} \bar{\Delta} s_{\Ab \Bb \Cb} \ ,
\end{equation}
which implies
\begin{equation}
       \delta s_{a b c} = - {\cal L}_{\bar{\xi}} s_{a b c} + \bar{\Psi}_a^{[\Ab} \bar{\Psi}_b^\Bb \bar{\Psi}_c^{\Cb]} 
       \bar{\Delta} s_{\Ab \Bb \Cb} \ .
\end{equation}
Now we can rewrite $\delta s^a$ as 
\begin{align}
   \delta s^a & = - \frac{1}{3!} \epsilon^{b c d a} {\cal L}_{\bar{\xi}} s_{b c d} + \frac{1}{3!} \epsilon^{b c d a} 
    \bar{\Psi}_{[b}^\Ab \bar{\Psi}_c^\Bb \bar{\Psi}_{d]}^\Cb \bar{\Delta} s_{\Ab \Bb \Cb} + \frac{1}{2} s^a g_{b c} \delta g^{b c} \cr
    & = - \left[\bar{\xi}^b \nabla_b s^a - \left(s^d \nabla_d \bar{\xi}^a - s^a \nabla_d \bar{\xi}^d\right)\right] 
    + \left(\frac{1}{3!} \epsilon^{b c d a} \bar{\Psi}_{[b}^\Ab \bar{\Psi}_c^\Bb \bar{\Psi}_{d]}^\Cb 
    \bar{\epsilon}_{\Ab \Bb \Cb}\right) \left(\frac{1}{3!} \bar{\epsilon}^{\Db \Eb \Fb} \bar{\Delta} s_{\Db \Eb \Fb}\right) \cr
    & + \frac{1}{2} s^a g_{b c} \delta g^{b c} 
     = - \left[\bar{\xi}^b \nabla_b s^a - \left(s^d \nabla_d \bar{\xi}^a - s^a \nabla_d \bar{\xi}^d\right)\right] 
    + \left(\frac{1}{3!} \bar{\epsilon}^{\Ab \Bb \Cb} \bar{\Delta} s_{\Ab \Bb \Cb}\right) \bu^a + \frac{1}{2} s^a g_{b c} \delta g^{b c} \ ,
\end{align} 
so that
\begin{align}
      \Theta_a \delta s^a & = - \left(2 s^b \nabla_{[b} \Theta_{a]} + \Gamma_s \Theta_a\right) \bar{\xi}^a  
                       - \frac{1}{3!} \Theta^{\Ab \Bb \Cb} \bar{\Delta} s_{\Ab \Bb \Cb} - \frac{1}{2} \Theta_c s^c g^{a b} \delta g_{a b} + {\cal BT} \ .
\end{align}
Like Eq.~\eqref{nconserv} for the matter component, when $\bar{\Delta} s_{\Ab \Bb \Cb} = 0$ we get the conservative part of the variation:
\begin{align}
      \left.\Theta_a \delta s^a\right|_{\cal C} & = - 2 s^b \nabla_{[b} \Theta_{a]} \bar{\xi}^a - \frac{1}{2} \Theta_c s^c g^{a b} \delta g_{a b} 
      + {\cal BT} \ .
\end{align}
However, when $\bar{\Delta} s_{\Ab \Bb \Cb} \neq 0$, the model is no longer ``non-dissipative'' since $s^a$ has a non-zero creation rate. The total 
entropy variation in the action principle can then be written as
\begin{align}
      \Theta_a \delta s^a & = \left.\Theta_a \delta s^a\right|_{\cal C} - \Gamma_\s \Theta_a \bar{\xi}^a 
      - \sum_{I = 1}^6 \frac{1}{3!} \left.\Theta^{\Ab \Bb \Cb} \bar{\Delta} s_{\Ab \Bb \Cb}\right|_I + {\cal BT} \ .
\end{align}

\subsection{External and Internal Index Structure of \texorpdfstring{$s_{\Ab \Bb \Cb}$}{The Entropy Space Form} and 
\texorpdfstring{$\bar{\Delta} s_{\Ab \Bb \Cb}$}{The Entropy Space Variation}}
\label{explicit_terms}

\noindent
The dissipation channels that show up in the equations of motion, $T^{a b}$, and $\Gamma_\s$, are given (albeit implicitly) in Eq.~\eqref{delbsabc}. Each 
term in $\{\xa,V^A,\bar{V}^\Ab,\gab,\bgab,\dgab\}$ makes a unique contribution to the dissipation. Evidently, the general problem has a rich structure. In order to make sense of this and build the relevant contributions, we need to explain how matter space degrees of freedom can be inserted into 
$s_{\Ab \Bb \Cb}$ to satisfy the index structure in Eq.~\eqref{delbsabc}.

Each contribution to $s_{\Ab \Bb \Cb}$ must have free indices $\{\Ab,\Bb,\Cb\}$ which are completely antisymmetric. Any other index 
carrying parts in each piece must appear in ``scalars'' (see \cite{andersson24:_ent1diss} for explicit constructions) in the sense that all indices 
must be summed over, whether it be particle space indices with particle space indices, entropy space indices with entropy space indices, or 
particle space indices with entropy space indices. For example, to include ${\cal S}_V$, the simplest choices are quadratic, such as $\ginvab V^A V^B$ and $\bginvab \bar{V}^\Ab \bar{V}^\Bb$. Note, however, that these combinations  necessarily involve degrees of freedom from ${\cal S}_g$. Another contribution associated with ${\cal S}_g$ is $\ginvab \bgab$. Finally, we can include ${\cal S}_{Lg}$ by having combinations like $\ginvab \dgab$, $\bginvab \dgab$, or $\ginvac \ginvbd \dgab \dgcd$. We will see later that these specific examples provide some guidance on imposing the second law of thermodynamics.

Each of the pieces in the Lagrangian variation of $s_{\Ab \Bb \Cb}$ can be represented in a more generic form 
$\bar{\Delta} s_{\Ab \Bb \Cb} = \left(\partial s_{\Ab \Bb \Cb}/\partial {\rm D_F}\right) \bar{\Delta} {\rm D_F}$, where $D_F$ is a degree of 
freedom taken from the set $\{{\cal S}_X,{\cal S}_V,{\cal S}_g,{\cal S}_{Lg}\}$. The term $\bar{\Delta} {\rm D_F}$ will have the same number of matter 
space indices as in the denominator of $\left(\partial s_{\Ab \Bb \Cb}/\partial {\rm D_F}\right)$ and these are summed over. But, as we have 
seen, the Lagrangian variations $\bar{\Delta} {\rm D_F}$ will have internal spacetime indices which are summed over. When 
$\Theta^{\Ab \Bb \Cb} \bar{\Delta} s_{\Ab \Bb \Cb}/3!$ is inserted into the action principle, there will necessarily be a gathering of terms with 
spacetime indices summed with $\xi^a$, others with $\bar{\xi}^a$, and some with $\delta g_{a b}$. How this works out will hopefully become clear as we progress.

We have found (and implemented) a way of writing the gathered terms using a  ``basis''built using the maps $\paa$, $\bpaa$, 
and the covariant derivative $\nabla^a$; specifically, a basis containing terms like $\paa$, $\bpaa$, $\paa \pbb$, $\paa \bpbb$, 
$\paa \nabla_c \pbb$, $\paa \nabla_c \bpbb$, $\dots$, $\bpbb \nabla_d \left(\nabla_c \paa\right)$. However, this collection of terms is not unique
because we could include the four-velocity, $u^a$, and the metric, $g^{a b}$. Working only with $\paa$, $\bpaa$, and 
$\nabla^a$ appears to provide a minimal set of terms involving somewhat  less ambiguity.

\section{Specific Variations of the Action}
\label{spevar}

\noindent
We have arrived at a point where the equations of motion, stress-energy-momentum tensor and entropy creation rate can be derived for a fairly generic model. We 
have shown that the terms $\sum_I \frac{1}{3!} \left.\Theta^{\Ab \Bb \Cb} \bar{\Delta} s_{\Ab \Bb \Cb}\right|_I$ contain $\bar{\xi}^a$, $\xi^a$, 
and $\delta g_{a b}$, meaning that the equations of motion, $T^{a b}$, and $\Gamma_\s$ will be modified depending on the specific choices we made. 
Specifically, using the formulas above for 
$\mu_a \delta n^a$ and $\Theta_a \delta s^a$, we find that the total variation of the action takes the form
\begin{align}
   \delta S_F & = \fint \left\{- \left(2 n^b \nabla_{[b} \mu_{a]}\right) \xi^a - \left(2 s^b \nabla_{[b} \Theta_{a]} + \Gamma_\s \Theta_a\right) 
    \bar{\xi}^a - \sum_{I = 1}^6 \frac{1}{3!} \left.\Theta^{\Ab \Bb \Cb} \bar{\Delta} s_{\Ab \Bb \Cb}\right|_I \right. \cr
      & \left. + \frac{1}{2} \left[\left(\Lambda - \mu_c n^c - \Theta_c s^c\right) g^{a b} + \mu^a n^b 
      + \Theta^a s^b\right] \delta g_{a b}\right\} + {\cal BT} \ .
\end{align}
In what follows, we will consider three different models based on an increasingly complex functional dependence of $s_{\Ab \Bb \Cb}$: 
\begin{align*}
    \text{(i)} \quad & \{\xa\}; \\
    \text{(ii)} \quad & \{\xa,V^A,\bar{V}^\Ab,\gab,\bgab\}; \\
    \text{(iii)} \quad & \{\xa,V^A,\bar{V}^\Ab,\gab,\bgab,\dgab\}. 
\end{align*}

\subsection{Model (i): Dependence of the Entropy on the Matter Space Coordinates \texorpdfstring{${\cal S}_X$}{Matter Space Coordinates}}
\label{sbxdxd}

\noindent
The simplest model (already explored in \cite{Andersson15:_dissfl_act}) we may consider is $s_{\Ab \Bb \Cb} =  s_{\Ab \Bb \Cb}\left({\cal S}_X\right)$, in which case the dissipative contribution comes from 
$\left.\bar{\Delta} s_{\Ab \Bb \Cb}\right|_1$. We have 
\begin{align}
   \delta S_F & = \fint \left\{- 2 n^b \nabla_{[b} \mu_{a]} \xi^a - \left(2 s^b \nabla_{[b} \Theta_{a]} + \Gamma_\s \Theta_a\right) 
    \bar{\xi}^a - \frac{1}{3!} \left.\Theta^{\Ab \Bb \Cb} \bar{\Delta} s_{\Ab \Bb \Cb}\right|_1 \right. \cr
      & \left. + \frac{1}{2} \left[\left(\Lambda - \mu_c n^c - \Theta_c s^c\right) g^{a b} + \mu^a n^b 
      + \Theta^a s^b\right] \delta g_{a b}\right\} + {\cal B.T.} \cr
      & = \fint \left\{- \left(2 n^b \nabla_{[b} \mu_{a]} - \left.R_a\right|_1\right) \xi^a - \left(2 s^b \nabla_{[b} \Theta_{a]} 
      + \Gamma_\s \Theta_a + \left.R_a\right|_1\right) \bar{\xi}^a \right. \cr
      & \left.+ \frac{1}{2} \left[\left(\Lambda - \mu_c n^c - \Theta_c s^c\right) g^{a b} + \mu^a n^b 
      + \Theta^a s^b\right] \delta g_{a b}\right\} + {\cal BT} \ ,
\end{align}
where
\begin{align}
          \left.R_a\right|_1 & = \frac{1}{3!} \Theta^{\Ab \Bb \Cb} \frac{\partial s_{\Ab \Bb \Cb}}{\partial \xd} \pda \ , \quad 
         u^a \left.R_a\right|_1 = 0 \label{res1} \ .
\end{align}
The equations of motion are
\begin{subequations}
\begin{align}
2 n^b \nabla_{[b} \mu_{a]} & = \left.R_a\right|_1 \ , \label{n1steom} \\
2 s^b \nabla_{[b} \Theta_{a]} + \Gamma_\s \Theta_a & = - \left.R_a\right|_1  \ ,\label{s1steom}
\end{align}
\end{subequations}
and the stress-energy-momentum tensor becomes
\begin{align}
T^{a b} & = \Psi g^{a b} + \mu^a n^b + \Theta^a s^b \ ,
\end{align}
with the generalized pressure $\Psi$  defined as
\begin{align}
\label{press}
    \Psi & = \Lambda - \mu_a n^a - \Theta_a s^a \ .
\end{align}

Note that, if we project the matter equation of motion along $n^a$ we find
\begin{align}
   0 & = n^a \left(2 n^b \nabla_{[b} \mu_{a]} - \left.R_a\right|_1\right) = - n u^a \left.R_a\right|_1 \ , \label{rperp}
\end{align}
which is consistent with Eq.~\eqref{res1}. If we add together Eqs.~\eqref{n1steom} and \eqref{s1steom} we find that
\begin{align}
   2 n^b \nabla_{[b} \mu_{a]} + 2 s^b \nabla_{[b} \Theta_{a]} + \Gamma_\s \Theta_a = 0 \ . \label{nps1steom}
\end{align} 
This is equivalent to $\nabla_b T^b{}_a = 0$. Further, if we project Eq.~\eqref{s1steom} along $s^b$ we arrive at a formula for $\Gamma_\s$; i.e.,
\begin{align}
         s^a \left(2 s^b \nabla_{[b} \Theta_{a]} + \Gamma_\s \Theta_a\right) & = - s^a \left.R_a\right|_1 \ ;
\label{gen_ent}
\end{align}
or,
\begin{align}
          \Gamma_\s & = \frac{1}{T_g} \bu^a \left.R_a\right|_1 \ . \label{1stentcre}
\end{align} 

It is fairly easy to show that the simple form of $s_{\Ab \Bb \Cb}$ examined here is consistent with
previous determinations of a general relativistic form of the Cattaneo equation \cite{Cattaneo48:_cateq} for heat flow; see, in particular, 
\cite{monsalvo11:_thesis}. We will take advantage of the identity $u^a \left.R_a\right|_1 = 0$ [\cf~Eq.~\eqref{res1}] by using it to write the 
entropy flux in terms of the particle frame of reference. The advantage is that $\left.R_a\right|_1$ is automatically perpendicular to the particle 
flux worldlines. 

Letting $\tilde{s}^a = \perp^a_b s^b$ represent the purely spatial part of the entropy flux with respect to the particle frame of reference and 
working out the projection leads to\footnote{Many previous works use the heat drift $\tilde{q}^a$ instead of the entropy drift $\tilde{s}^a$; the two 
are related via $\tilde{q}^a = T^* \tilde{s}^a$.}
\begin{align}
      s^a & = s^* u^a + \tilde{s}^a \ , \quad 
      s^* = - u_b s^b \ , \quad
      \tilde{s}^a = \perp^a_b s^b \ . 
      \label{entdecomp}
\end{align} 
The particle momentum $\mu_a$ can be decomposed by direct substitution into it the above relation for $s^a$; i.e.,
\begin{align}
      \mu_a & = \mu^* u^a + \Asn \tilde{s}^a \ , \quad 
      \mu^* = - u^b \mu_b = \Bn n + \Asn s^* \ . 
      \label{partmom}
\end{align} 
For the decomposition of the thermal four-momentum $\Theta_a$, we use again particle frame temporal and spatial parts:
\begin{align}
      \Theta_a & = T^* u_a + \tilde{\Theta}_a \ , \quad 
      T^* = - u^a \Theta_a \ , \quad
      \tilde{\Theta}_a = \perp^b_a \Theta_b \ . 
      \label{entmom}
\end{align}

For later convenience, we may use the entropy flux as perceived in the particle frame; i.e.,~Eq.~\eqref{entdecomp}, to introduce a relative drift velocity that should be small in the near-equilibrium regime. Retaining only linear pieces in the 
entropy drift, we have
\begin{align}
      \bu^a & \approx u^a + \tilde{w}^a \ ,
      \label{lindrift}
\end{align} 
and it follows that 
\begin{align}
      s^2 & \equiv - g_{a b} \left(s^* u^a + \tilde{s}^a\right) \left(s^* u^a + \tilde{s}^a\right) = \left(s^*\right)^2 \left(1 - \left|\tilde{w}^a\right|^2\right) 
                \approx \left(s^*\right)^2 \ , \quad 
\end{align} 
which implies $s \approx s^*$. 
We will make explicit use of this approximation later.

If we substitute the decomposition of $s^a$ into Eq.~\eqref{entmom}, then we get
\begin{align}
    \Theta_a & = \Bs \left(s^* u_a + \tilde{s}_a\right) + \Asn n u_a = \left(\Bs s^* + \Asn n\right) u_a + \Bs \tilde{s}_a \ , 
\end{align}
which implies
\begin{align}
    T^* & = \Bs s^* + \Asn n \ , \quad \tilde{\Theta}_a = \Bs \tilde{s}_a \ .
\end{align}
This leads to the final form
\begin{align}
    \Theta_a & = T^* u_a + \Bs \tilde{s}_a \ . \label{thetadecomp}
\end{align}

We already have an expression for the entropy creation rate from Eq.~\eqref{1stentcre}. To get at a different form, necessary for deriving the Cattaneo 
Equation, it is useful to project the quantity $2 s^b \nabla_{[b} \Theta_{a]}$ along $u^a$ to find
\begin{align}
   2 u^a s^b \nabla_{[b} \Theta_{a]} & = - \tilde{s}^a \left(\nabla_a T^* + T^* a_a + \dot{\cal B}^\s \tilde{s}_a 
             + 2 \Bs u^b \nabla_{[b} \tilde{s}_{a]}\right) \ ,
\end{align}
where a dot indicates a proper time derivative from the perspective of $u^a$ (i.e.,  $\dot{O} = u^a \nabla_a O$). Now if we take Eq.~\eqref{s1steom} and project its free index along $u^a$ we find the second 
form of the entropy rate:
\begin{align}
     T^* \Gamma_\s & = - \tilde{s}^a \left(\nabla_a T^* + T^* a_a + \dot{\cal B}^\s \tilde{s}_a + 2 \Bs u^b \nabla_{[b} \tilde{s}_{a]}\right) \ . 
     \label{2stentcre}
\end{align}
At this point, we invoke the second law of thermodynamics  to impose the constitutive relation (with $\lambda \geq 0$)
\begin{align}
    \tilde{s}_a & = - \frac{\lambda}{T^*} \perp^c_a \left(\nabla_c T^* + T^* a_c + \dot{\cal B}^\s \tilde{s}_c 
                           + 2 \Bs u^b \nabla_{[b} \tilde{s}_{c]}\right) \ , 
    \label{entconstitutive}
\end{align} 
which leads to the positive-definite form
\begin{align}
     \Gamma_\s & = \frac{\tilde{s}_a \tilde{s}^a}{\lambda} \geq 0 \ . \label{1stentcreb}
\end{align}

With the above relation in hand,  we refocus the meaning of the constitutive relation in Eq.~\eqref{entconstitutive} and argue that it  serves as an 
equation for the heat flux. That is, we solve for the last term to produce the differential equation 
\cite{eckart40:_rel_diss_fluid,carter76,stew77,Israel79:_kintheo1,Monsalvo10:_thermogr,monsalvo11:_thesis}  
\begin{align}
    \frac{2 \lambda \Bs}{\left(T^* + \lambda \dot{\cal B}^\s\right)} u^b \nabla_{[b} \tilde{s}_{a]} + \tilde{s}_a & = 
    - \frac{\lambda}{T^* + \lambda \dot{\cal B}^\s} \perp^b_a \left(\nabla_b T^* + T^* a_b\right) \ . \label{tele_eqn}
\end{align} 
This is the Cattaneo equation which leads to causal flow for $\tilde{s}_a$ with relaxation timescales and heat conductivity given by 
\begin{equation}
    \tau = \frac{2 \lambda \Bs}{T^* + \lambda \dot{\cal B}^\s} \;,\quad \kappa = \frac{\lambda}{T^* + \lambda \dot{\cal B}^\s} \;.
\end{equation}

Now we use the constitutive relation to find a form for $\left.R_a\right|_1$ in terms of the field variables. The main constraint comes from 
equating Eq.~\eqref{1stentcre} to Eq.~\eqref{1stentcreb}, which leads to 
\begin{align}
    \tilde{s}^a \left(\left.R_a\right|_1 - \frac{s T_g}{\lambda} \tilde{s}_a\right) & = 0 \ . 
    \label{ra1eq}
\end{align}
To linear order in $\tilde{s}_a$, which is what we will need later, this is satisfied by
\begin{align}
    \left.R_a\right|_1 & = \frac{s T_g}{\lambda} \tilde{s}_a + \chi \tilde{\epsilon}_{a b c} \tilde{s}^b \left.R^c\right|_1  
    \ , \quad \tilde{\epsilon}_{a b c} = u^d \epsilon_{d a b c} \ .
    \label{ra1sol}
\end{align}
Since the leading order term in $\left.R_a\right|_1$ is $\tilde{s}^a$, that means that the last term in Eq.~\eqref{ra1sol} is (at least) of second order in 
the entropy drift, and can therefore be neglected. The final result is then
\begin{align}
    \left.R^a\right|_1 & \approx \frac{s T_g}{\lambda} \tilde{s}^a \ .
    \label{res1asol}
\end{align}

Finally, in  \cite{andersson24:_ent1diss}, a dissipative single-fluid model of particles and entropy was constructed. In that case the action principle began 
with only one four-velocity and two constituents. This was achieved by assuming that two matter spaces exist but that they are related to each 
other via a diffeomorphism. This construction resulted in a single field equation. The difference here is that we are starting with a system 
that has two dynamically independent fluxes.

For the simple case considered in this section, it is straightforward to see how an effective single-fluid description emerges. From \cref{entconstitutive,1stentcreb} we have $\Gamma_s \propto \lambda$, while the entropy drift is itself proportional to $\lambda$. Thus, as equilibrium is approached ($\lambda \to 0$), both the entropy production and the relative entropy flow are suppressed, and the entropy flux becomes effectively comoving with the matter sector. Since it is also the case that $\tau \propto \lambda$, this regime may be interpreted as the near-equilibrium (or late-time) behaviour of the underlying non-linear model. This is notably different from the parabolic limit of the Cattaneo equation, which rather corresponds to an asymptotic reduction valid when the observation timescale is much larger than the relaxation timescale.
This means $\left.R^a\right|_1$ is also driven to zero, but what we end up with is a double-constituent, 
single-fluid system of the type described in Sec.~\ref{zerodissfluids}: There are two equations, one of which can be thought of as an equation 
of motion for $u^a$ and the other serves as a constraint equation for $T_g$ (as in the case of Eq.~\eqref{tolman}).  

\subsection{Model (ii): Dependence of the Entropy on the Matter Space Coordinates \texorpdfstring{${\cal S}_X$}{Matter Space Coordinates}, the 
Velocities \texorpdfstring{${\cal S}_V$}{Matter Space Velocities}, and Metrics \texorpdfstring{${\cal S}_g$}{Matter Space Metrics}}
\label{3metricsmod}

\noindent
Stepping up the complexity, let us consider a model where the entropy three form depends on ${\cal S}_X$, ${\cal S}_V$ and ${\cal S}_g$. As we will see there is, strictly speaking, no way for this model to make the entropy creation rate positive definite for each of the terms, because there are no 
time derivatives of the matter space variables in $s_{\Ab \Bb \Cb}$ and hence no way for the resistive and dissipative coefficients to generate 
terms inside themselves like $\theta$ and $\nabla_a u^b$ (\cf~the discussion in Sec.~\ref{uafib}). 
This problem will be addressed with the next model, which includes
 time derivatives of matter space variables. However, we will nevertheless  proceed with the analysis for two reasons: (i) Several new resisitive vectors and dissipation tensors will emerge  (also in the next 
section) and it is useful to introduce them now to establish identities they satisfy and introduce some notational rules; and (ii) we will show that the Cattaneo 
equation gets modified via the inclusion of resistivity but not dissipation, an outcome that will be repeated in the next section.

The obvious next step would be to consider the dependence on the matter space velocities ${\cal S}_V$. But it is clear from the discussion in Sec.~\ref{explicit_terms} 
that the matter space metrics are necessary when the velocities are included. Hence, the entropy space three-form is given by 
$s_{\Ab \Bb \Cb} =  s_{\Ab \Bb \Cb}\left({\cal S}_X,{\cal S}_V,{\cal S}_g\right)$, with the dissipative 
contributions coming from $\left.\bar{\Delta} s_{\Ab \Bb \Cb}\right|_1$, $\left.\bar{\Delta} s_{\Ab \Bb \Cb}\right|_2$, 
$\left.\bar{\Delta} s_{\Ab \Bb \Cb}\right|_3$, $\left.\bar{\Delta} s_{\Ab \Bb \Cb}\right|_4$, and $\left.\bar{\Delta} s_{\Ab \Bb \Cb}\right|_5$. The 
variation of the action now looks like  
\begin{align}
   \delta S_F & = \fint \left\{- \left(2 n^b \nabla_{[b} \mu_{a]}\right) \xi^a - \left(2 s^b \nabla_{[b} \Theta_{a]} + \Gamma_\s \Theta_a\right) 
     \bar{\xi}^a - \frac{1}{3!} \left.\Theta^{\Ab \Bb \Cb} \bar{\Delta} s_{\Ab \Bb \Cb}\right|_1 \right. \cr
     & \left. - \frac{1}{3!} \left.\Theta^{\Ab \Bb \Cb} \bar{\Delta} s_{\Ab \Bb \Cb}\right|_2 - \frac{1}{3!} \left.\Theta^{\Ab \Bb \Cb} \bar{\Delta} 
      s_{\Ab \Bb \Cb}\right|_3 - \frac{1}{3!} \left.\Theta^{\Ab \Bb \Cb} \bar{\Delta} s_{\Ab \Bb \Cb}\right|_4 - \frac{1}{3!} \left.\Theta^{\Ab \Bb \Cb} 
      \bar{\Delta} s_{\Ab \Bb \Cb}\right|_5 \right. \cr
      & \left. + \frac{1}{2} \left[\left(\Lambda - \mu_c n^c  - \Theta_c s^c\right) g^{a b} + \mu^a n^b + \Theta^a s^b\right] \delta g_{a b}\right\}
      + {\cal B.T.} \cr
      & = \fint \left\{- \left[2 n^b \nabla_{[b} \mu_{a]} - \left.R_a\right|_1 - \left.D_{a b}\right|_2 \bu^b - \left.R_b\right|_3 \nabla_a u^b -
       2 \left.D_{b a c}\right|_5 g^{b c} \right. \right. \cr
       & \left. \left. - \nabla^b \left(\perp^c_a \left.R_c\right|_3 u_b - \left.R_a\right|_2 \bu_b - 2 \left.D_{a b}\right|_5\right)\right] \xi^a 
      - \left[2 s^b \nabla_{[b} \Theta_{a]} + \Gamma_\s \Theta_a + \left.R_a\right|_1 + \left.D_{a b}\right|_2 \bu^b \right. \right. \cr 
      & \left. \left. + \left.R_b\right|_3 \nabla_a u^b + 2 \left.D_{b a c}\right|_5 g^{b c} + \nabla^b \left(\left.R_a\right|_3 u_b 
      - \bar{\perp}^c_a \left.R_c\right|_2 \bu_b + 2 \left.D_{a b}\right|_4\right)\right] \bar{\xi}^a \right. \cr
      & \left. + \frac{1}{2} \left(\Psi g^{a b} + \mu^a n^b + \Theta^a s^b + 2 \left.D^{a b}\right|_4 + 2 \left.D^{a b}\right|_5 
      - \bu^c \left.R_c\right|_2 \bu^a \bu^b - u^c \left.R_c\right|_3 u^a u^b\right) \delta g_{a b}\right\} \cr
      & + {\cal BT} \ ,
\end{align}
where
\begin{subequations}
\begin{align}
         \left.R_a\right|_2 & = \frac{1}{3!} \Theta^{\Ab \Bb \Cb} \frac{\partial s_{\Ab \Bb \Cb}}{\partial V^\D} \pda \ , \quad 
         u^a \left.R_a\right|_2 = 0 \label{res2} \ , \\
         \left.R_a\right|_3 & = \frac{1}{3!} \Theta^{\Ab \Bb \Cb} \frac{\partial s_{\Ab \Bb \Cb}}{\partial \bar{V}^\Db} \bpda \ , \quad 
         \bu^a \left.R_a\right|_3 = 0 \label{res3} \ , \\ 
         \left.D_{a b}\right|_2 & = \frac{1}{3!} \Theta^{\Ab \Bb \Cb} \frac{\partial s_{\Ab \Bb \Cb}}{\partial V^\D} \nabla_a \pdb 
         = \left.D_{b a}\right|_2 \ , \quad 
         u^b \left.D_{a b}\right|_2 = - \left.R_b\right|_2 \nabla_a u^b \label{diss2} \ , \\
         \left.D_{a b}\right|_4 & = \frac{1}{3!} \Theta^{\Ab \Bb \Cb} \frac{\partial s_{\Ab \Bb \Cb}}{\partial \bgde} \bpda \bpeb 
         = \left.D_{b a}\right|_4 \ , \quad 
         \bu^b \left.D_{a b}\right|_4 = \bu^b \left.D_{b a}\right|_4 = 0 \label{diss4} \ , \\ 
         \left.D_{a b}\right|_5 & = \frac{1}{3!} \Theta^{\Ab \Bb \Cb} \frac{\partial s_{\Ab \Bb \Cb}}{\partial \gde} \pda \peb 
         = \left.D_{b a}\right|_5 \ , \quad 
         u^b \left.D_{a b}\right|_5 = u^b \left.D_{b a}\right|_5 = 0 \label{diss5a} \ , \\ 
         \left.D_{a c b}\right|_5 & = \frac{1}{3!} \Theta^{\Ab \Bb \Cb} \frac{\partial s_{\Ab \Bb \Cb}}{\partial \gde} \pda \nabla_c \peb 
         =  \left.D_{a b c}\right|_5 \ , \quad 
         u^a \left.D_{a c b}\right|_5 = 0 \ , \quad
         u^c \left.D_{a c b}\right|_5 = - \left.D_{a c}\right|_5 \nabla_b u^c \label{diss5b} \ .                
\end{align}
\end{subequations}
The equations of motion are
\begin{subequations}
\begin{align}
2 n^b \nabla_{[b} \mu_{a]} & = \left.R_a\right|_1 + \left.D_{a b}\right|_2 \bu^b + \left.R_b\right|_3 \nabla_a u^b +
       2 \left.D_{b a c}\right|_5 g^{b c}  \cr
       & + \nabla^b \left(\perp^c_a \left.R_c\right|_3 u_b - \left.R_a\right|_2 \bu_b - 2 \left.D_{a b}\right|_5\right) \ , \\
2 s^b \nabla_{[b} \Theta_{a]} + \Gamma_\s \Theta_a & = - \left.R_a\right|_1 - \left.D_{a b}\right|_2 \bu^b 
       - \left.R_b\right|_3 \nabla_a u^b - 2 \left.D_{b a c}\right|_5 g^{b c} \cr
       & + \nabla^b \left(\bar{\perp}^c_a \left.R_c\right|_2 \bu_b - \left.R_a\right|_3 u_b - 2 \left.D_{a b}\right|_4\right) \ ,
\end{align}
\end{subequations}
and the stress-energy-momentum tensor is
\begin{align}
   T^{a b} & = \Psi g^{a b} + \mu^a n^b + \Theta^a s^b + 2 \left.D^{a b}\right|_4 + 2 \left.D^{a b}\right|_5 - \bu^c \left.R_c\right|_2 \bu^a \bu^b 
                      - u^c \left.R_c\right|_3 u^a u^b \ ,
\end{align}
with the generalized pressure $\Psi$ taking the same form as before. 

A projection of the particle equation of motion along $u^a$ still leads to zero on the left hand side. The same projection on the right hand side can 
be shown to equal zero identically by making use of the projections  in Eqs.~\eqref{res2}, \eqref{res3}, and \eqref{diss2}. As in the 
previous model, the entropy creation rate can be found by projecting the entropy equation of motion along $\bu^a$. By again making use of the 
projections given in Eqs.~\eqref{res2}, \eqref{res3}, and \eqref{diss2}, we find that $\Gamma_\s$ is given by
\begin{align}
     T_g \Gamma_\s & = \left.R_a\right|_1 \bu^a + \left.R_a\right|_2 \bar{a}^a + \left.R_a\right|_3 \left(\bu^b \nabla_b u^a - u^b \nabla_b \bu^a\right) 
           + \left.D_{a b}\right|_2 \bu^a \bu^b - 2 \left.D_{a b}\right|_4 \nabla^{(a} \bu^{b)} \cr
       & + 2 \left.D_{b a c}\right|_5 g^{b c} \bu^a \ .
\end{align}

Here we note the appearance of three accelerations: 1) $\bar{a}^a$, the acceleration of the entropy fluid defined on spacetime and not with respect 
to any particular frame; 2) $\bu^b \nabla_b u^a $, which can be interpreted as the acceleration of the particles with respect to the entropy fluid; 
and 3) $u^b \nabla_b \bu^a$, the acceleration of the entropy fluid with respect to the particles. Earlier, we noted the introduction of acceleration  
in the Cattaneo equation which determines the time evolution of the entropy flux with respect to the particles. But here, the acceleration enters 
already at the level of the entropy creation rate as defined with respect to spacetime and not a particular frame. 

Let us examine the impact of the new dissipation channels on the Cattaneo equation. Even though the second law is not 
guaranteed, it is useful to see how the nature of the Cattaneo equation is affected and if there is still an argument that $\tilde{s}^a$ 
will be driven to zero at late times. This analysis will be useful when the single-fluid limit is discussed in the next section.  

A main step towards extracting the Cattaneo Equation is to project into the local spacelike sections perpendicular to $u^a$ at each point along the 
worldlines. We already  know that $\left.R_a\right|_1$, $\left.R_a\right|_2$, and $\left.D_{a b}\right|_5$ live completely in the spacelike sections 
and that $\left.D_{a c b}\right|_5$ is spacelike on the first index. It is worthwhile to decompose $\left.R_b\right|_3$, $\left.D_{a b}\right|_2$, and 
$\left.D_{a b}\right|_4$ 
into pieces parallel and normal to $u^a$; i.e.,~we can show that
\begin{align}
         \left.R_a\right|_3 & =  \left(\frac{1}{s^*} \tilde{s}^b \left.\tilde{R}_b\right|_3\right) u_a + \left.\tilde{R}_a\right|_3 \ , \quad 
         \left.\tilde{R}_a\right|_3 = \perp^b_a \left.R_b\right|_3 \ , 
\end{align}
where we have used
\begin{align}
         \bu^a \left.R_a\right|_3 = 0 \quad \Longrightarrow \quad u^a \left.R_a\right|_3 = - \frac{1}{s^*} \tilde{s}^a \left.\tilde{R}_a\right|_3 \ ;
\end{align}
and then we have for the dissipation tensors 
\begin{subequations}
\begin{align}
     \left.D_{a b}\right|_2 & =  \left.\tilde{D}\right|_2 u_a u_b + 2  \left.\tilde{D}_{(a}\right|_2 u_{b)} + \left.\tilde{D}_{a b}\right|_2 \ , \\
     \left.\tilde{D}\right|_2 & = u^a u^b \left.D_{a b}\right|_2 = - \left.R_a\right|_2 a^a \ , \\
     \left.\tilde{D}_a\right|_2 & = - u^c \perp^b_a \left.D_{b c}\right|_2 = \left(\left.R_b\right|_2 a^b\right) u_a + \left.R_b\right|_2 \nabla_a u^b \ ,\\
     \left.\tilde{D}_{a b}\right|_2 & = \perp^c_a \perp^d_b \left.D_{c d}\right|_2 \ ,                  
\end{align}
\end{subequations}
and
\begin{subequations}
\begin{align}
         \left.D_{a b}\right|_4 & =  \left.\tilde{D}\right|_4 u_a u_b + 2  \left.\tilde{D}_{(a}\right|_4 u_{b)} + \left.\tilde{D}_{a b}\right|_4 \ , \\
         \left.\tilde{D}\right|_4 & = u^a u^b \left.D_{a b}\right|_4 = \frac{1}{\left(s^*\right)^2} \tilde{s}^a \tilde{s}^b \left.\tilde{D}_{a b}\right|_4 \ , \\
         \left.\tilde{D}_a\right|_4 & = - u^c \perp^b_a \left.D_{b c}\right|_4 = \frac{1}{s^*} \perp^b_a \tilde{s}^c \left.\tilde{D}_{b c}\right|_4 \ ,\\
         \left.\tilde{D}_{a b}\right|_4 & = \perp^c_a \perp^d_b \left.D_{c d}\right|_4 \ , 
         \label{dab4decomp}                 
\end{align}
\end{subequations}
where we have used
\begin{align}
         \bu^b \left.D_{a b}\right|_4 & = 0 \quad \Longrightarrow \quad 
         u^b \left.D_{a b}\right|_4 = - \frac{1}{s^*} \tilde{s}^b \left.D_{a b}\right|_4 \ .                         
\end{align} 
We see that all the components of $\left.D_{a b}\right|_4$ have been reduced to a dependence on only the purely spatial (as defined by $u^a$) 
pieces $\left.\tilde{D}_{a b}\right|_4$. These are the stress-carrying parts of $\left.D_{a b}\right|_4$. These stresses also manifest themselves as 
an energy $\left.\tilde{D}\right|_4$ and a momentum $\left.\tilde{D}_a\right|_4$. Because $\left.D_{a b}\right|_5$ is perpendicular to $u^a$ on 
both indices, its decomposition is simply the stress-carrying
\begin{align}
        \left.\tilde{D}_{a b}\right|_5 = \perp^c_a \perp^d_b \left.D_{c d}\right|_5 = \left.D_{a b}\right|_5 \ .
\end{align}
A final thing we can do is erect mutually orthogonal basis vectors to decompose 
$\left.R_a\right|_2$ and $\left.\tilde{R}_a\right|_3$; namely,
\begin{subequations}
\begin{align}
         \left.\tilde{R}_a\right|_2 & = \left.r_1\right|_2 \tilde{s}_a + \left.r_2\right|_2 \left(\perp^b_a - \tilde{s}_a \tilde{s}^b/\left|\tilde{s}\right|^2\right) 
         \left.\tilde{R}_b\right|_2 + \left.r_3\right|_2 \epsilon_{a b c} \tilde{s}^b \left.\tilde{R}^c\right|_2 \ , \label{r2decomp} \\
         \left.\tilde{R}_a\right|_3 & = \left.r_1\right|_3 \tilde{s}_a + \left.r_2\right|_3 \left(\perp^b_a - \tilde{s}_a \tilde{s}^b/\left|\tilde{s}\right|^2\right) 
         \left.\tilde{R}_b\right|_3 + \left.r_3\right|_3 \epsilon_{a b c} \tilde{s}^b \left.\tilde{R}^c\right|_3 \ . \label{r3decomp}
\end{align}
\end{subequations}

Following the same process as in the case of the previous model, we find that the entropy creation rate is given by
\begin{align}
T^* \Gamma_\s & = - \tilde{s}^a \left\{2 \Bs u^c \nabla_{[c} \tilde{s}_{a]} + \nabla_a T^* 
     + \left(T^* - \frac{s^*}{s^3} \left.r_1\right|_2 \left|\tilde{s}\right|^2\right) a_a \right. \cr
    & \left. + \left(\dot{\cal B}^\s + \nabla^c \left[\frac{1}{s^*} \left(\left.r_1\right|_3 
    - \left(\frac{s^*}{s}\right)^3 \left.r_1\right|_2\right) u_c - \frac{s^*}{s^3} \left.r_1\right|_2 \tilde{s}_c\right]\right) \tilde{s}_a \right. \cr
    & \left. - \frac{2 s^*}{s^3} \left.r_1\right|_2 \tilde{s}^c \nabla_{(c} \tilde{s}_{a)} - \frac{\left|\tilde{s}\right|^2}{s^3} \left.r_1\right|_2 
    \tilde{s}^c \nabla_{(c} u_{a)}\right\} \cr 
    & - 2 \nabla_a \left(\left.\tilde{D}\right|_4 u^a + \left.\tilde{D}^a\right|_4\right) 
    - 2 \left(\left.\tilde{D}_{a b}\right|_4 + \left.\tilde{D}_{a b}\right|_5\right) \nabla^{(a} u^{b)} \ . 
     \label{4thentcre}
\end{align}
Moreover, we can make the first term positive-definite if we set
\begin{align}
     \tilde{s}_a & = - \frac{\lambda}{T^*} \perp^b_a \left\{2 \Bs u^c \nabla_{[c} \tilde{s}_{b]} + \nabla_b T^* 
     + \left(T^* - \frac{s^*}{s^3} \left.r_1\right|_2 \left|\tilde{s}\right|^2\right) a_b \right. \cr
    & \left. + \left(\dot{\cal B}^\s + \nabla^c \left[\frac{1}{s^*} \left(\left.r_1\right|_3 
    - \left(\frac{s^*}{s}\right)^3 \left.r_1\right|_2\right) u_c - \frac{s^*}{s^3} \left.r_1\right|_2 \tilde{s}_c\right]\right) \tilde{s}_b \right. \cr
    & \left. - \frac{2 s^*}{s^3} \left.r_1\right|_2 \tilde{s}^c \nabla_{(c} \tilde{s}_{b)} - \frac{\left|\tilde{s}\right|^2}{s^3} \left.r_1\right|_2 
    \tilde{s}^c \nabla_{(c} u_{b)}\right\} \ ,
\end{align}
from which we infer 
\begin{align}
   2 \tau_2 u^b \nabla_{[b} \tilde{s}_{a]} & + \tilde{s}_a = - \frac{\tau_2}{\Bs} \perp^b_a \left[\nabla_b T^* + \left(T^* - \frac{s^*}{s^3} 
   \left.r_1\right|_2 \left|\tilde{s}\right|^2\right) a_b - \frac{2 s^*}{s^3} \left.r_1\right|_2 \tilde{s}^c \nabla_{(c} \tilde{s}_{b)} \right. \cr
   & \left. - \frac{\left|\tilde{s}\right|^2}{s^3} \left.r_1\right|_2 
    \tilde{s}^c \nabla_{(c} u_{b)}\right] \ ,
\end{align}
where
\begin{align}
     \tau_2 & = \lambda \Bs\left/\left(T^* + \lambda \dot{\cal B}^\s + \lambda \nabla^c \left[\frac{1}{s^*} \left(\left.r_1\right|_3 
    - \left(\frac{s^*}{s}\right)^3 \left.r_1\right|_2\right) u_c - \frac{s^*}{s^3} \left.r_1\right|_2 \tilde{s}_c\right]\right)\right. \ .
\end{align}
In this equation we recognize some of the features from the simple Cattaneo equation. For example, at late times $\tau_2$ it can be expected that $\tilde{s}^a \to 0$. However, there are not enough 
matter space degrees of freedom to make $\Gamma_\s \geq 0$ in Eq.~\eqref{4thentcre}, since the only matter space metric terms that can be 
envisioned are those like $\ginvab \bgab$ and $\bginvab \gab$, which cannot deliver derivatives.  In fact, the only move we can make towards a positive entropy 
creation rate is to set $\left.\tilde{D}_{a b}\right|_4 = - \left.\tilde{D}_{a b}\right|_5$, in which case $\Gamma_\s \to 0$.

Even though the entropy creation rate is not guaranteed to be positive definite, we  have a new, more complicated version of the Cattaneo equation which implies the entropy 
flux with respect to the particles, $\tilde{s}_a$, will be driven to zero. This means that the remaining entropy flow is parallel to that of the 
particles---$\bu^a = u^a$. We now use the ``entropy drift velocity'' from \eqref{lindrift}, i.e., $\tilde{w}_a = \tilde{s}_a / s^*$. From the decompositions of $\left.\tilde{R}_a\right|_2$ and $\left.\tilde{R}_a\right|_3$, i.e.,~Eqs.~\eqref{r2decomp} and 
\eqref{r3decomp}, respectively, we can show now that $\left.\tilde{R}_a\right|_2 = \left.\tilde{R}_a\right|_3 \to 0$ as $\tilde{w}_a \to 0$.

Earlier, we constructed the constitutive solution Eq.~\eqref{ra1sol} to Eq.~\eqref{ra1eq} for the resistivity $\left.R_a\right|_1$. This is relevant 
here because, assuming the entropy drift is small allows us to write the additional resistivities as
\begin{subequations}
\begin{align}
         \left.\tilde{R}_a\right|_2 & \approx \frac{s \left.r_1\right|_2}{1 - \left.r_2\right|_2} \tilde{w}_a 
                              + \frac{s \left.r_3\right|_2}{1 - \left.r_2\right|_2} \epsilon_{a b c} \tilde{w}^b \left.\tilde{R}^c\right|_2 \ , \label{r2approx} \\
         \left.\tilde{R}_a\right|_3 & \approx \frac{s \left.r_1\right|_3}{1 - \left.r_2\right|_3} \tilde{w}_a 
                      + \frac{s \left.r_3\right|_3}{1 - \left.r_2\right|_3} \epsilon_{a b c} \tilde{w}^b \left.\tilde{R}^c\right|_3 \ . \label{r3approx}
\end{align}
\end{subequations}
Both of these have the same form as in Eq.~\eqref{ra1sol}, and so therefore, following the logic that leads to Eq.~\eqref{res1asol}, we can write (to first order in the entropy drift)
\begin{align}
         \left.\tilde{R}_a\right|_2 & \approx \frac{s \left.r_1\right|_2}{1 - \left.r_2\right|_2} \tilde{w}_a \ , \quad
         \left.\tilde{R}_a\right|_3  \approx \frac{s \left.r_1\right|_3}{1 - \left.r_2\right|_3} \tilde{w}_a \ . 
\end{align}

The single fluid limit results, again, in two field equations, taking the form 
\begin{subequations}
\begin{align}
2 n^b \nabla_{[b} \mu_{a]} & + 2 s^b \nabla_{[b} \Theta_{a]} + \Gamma_\s \Theta_a  = - 2 \nabla^b \left(\left.D_{a b}\right|_4 
             + \left.D_{a b}\right|_5\right) \ , \\
2 s^b \nabla_{[b} \Theta_{a]} & + \Gamma_\s \Theta_a = - 2 \left.D_{b a c}\right|_5 g^{b c} - 2 \nabla^b \left.D_{a b}\right|_4 \ .
\end{align}
\end{subequations}
Projecting the entropy flux equation of motion along $u^a$ we get the single fluid limit for the entropy creation rate ($T^* = T$); i.e.,
\begin{align}
T_g \Gamma_\s & = - 2 \left(\left.\tilde{D}_{a b}\right|_4 + \left.\tilde{D}_{a b}\right|_5\right) \nabla^{(a} u^{b)} \ . 
\end{align}
Projecting the same equation of motion along $\perp^b_a$ we get
\begin{align}
     s \left(T_g a_b + \perp^b_a \nabla_b T_g\right) & = - 2 \left(\nabla^b \left.\tilde{D}_{a b}\right|_4  + \left.D_{b a c}\right|_5 g^{b c}\right) \ .
\end{align}

Again, we see that the form of the temperature profile is complicated by the 
presence of the dissipation tensors. Of course, as stressed before, this model is seriously flawed thermodynamically, so the complications could be due to 
this. However, we will note this appearance of dissipation tensors again in the next model, the major difference being that that model has 
enough matter degrees of freedom that the second law can be satisfied. 

\subsection{Model (iii): Dependence of the Entropy on the Matter Space Coordinates \texorpdfstring{${\cal S}_X$}{Matter Space Coordinates}, the Matter 
Space Velocities \texorpdfstring{${\cal S}_V$}{Matter Space Velocities}, the Matter Space Metrics 
\texorpdfstring{${\cal S}_g$}{Matter Space Metrics}, and Matter Space Metric Derivatives \texorpdfstring{${\cal S}_{Lg}$}{Matter Space Metric 
Derivatives}}
\label{metricdermod}

\noindent
For the final model we consider the case when $s_{\Ab \Bb \Cb} =  s_{\Ab \Bb \Cb}\left({\cal S}_X,{\cal S}_V,{\cal S}_g,{\cal S}_{Lg}\right)$. This leads to
\begin{align}
   \delta S_F & = \fint \left\{- \left(2 n^b \nabla_{[b} \mu_{a]}\right) \xi^a - \left(2 s^b \nabla_{[b} \Theta_{a]} + \Gamma_\s \Theta_a\right) 
     \bar{\xi}^a - \frac{1}{3!} \left.\Theta^{\Ab \Bb \Cb} \bar{\Delta} s_{\Ab \Bb \Cb}\right|_1 \right. \cr
     & \left. - \frac{1}{3!} \left.\Theta^{\Ab \Bb \Cb} \bar{\Delta} s_{\Ab \Bb \Cb}\right|_2 - \frac{1}{3!} \left.\Theta^{\Ab \Bb \Cb} \bar{\Delta} 
      s_{\Ab \Bb \Cb}\right|_3 - \frac{1}{3!} \left.\Theta^{\Ab \Bb \Cb} \bar{\Delta} s_{\Ab \Bb \Cb}\right|_4 - \frac{1}{3!} \left.\Theta^{\Ab \Bb \Cb} 
      \bar{\Delta} s_{\Ab \Bb \Cb}\right|_5 \right. \cr
      & \left. - \frac{1}{3!} \left.\Theta^{\Ab \Bb \Cb} \bar{\Delta} s_{\Ab \Bb \Cb}\right|_6 + \frac{1}{2} \left[\left(\Lambda - \mu_c n^c  
      - \Theta_c s^c\right) g^{a b} + \mu^a n^b + \Theta^a s^b\right] \delta g_{a b}\right\} + {\cal B.T.} \cr
      & = \fint \left\{- \left[2 n^b \nabla_{[b} \mu_{a]} - \left.R_a\right|_1 - \left.D_{a b}\right|_2 \bu^b - \left.R_b\right|_3 \nabla_a u^b -
       2 \left.D_{b a c}\right|_5 g^{b c} \right. \right. \cr
       & \left. \left. - \nabla^b \left(\perp^c_a \left.R_c\right|_3 u_b - \left.R_a\right|_2 \bu_b - 2 \left.D_{a b}\right|_5 
       - 2 \left.D_{c d}\right|_6 u_a u_b \nabla^{(c} u^{d)} + 2 \nabla_c \left(u^c \left.D_{a b}\right|_6\right) \right. \right. \right. \cr
       & \left. \left. \left. + 2 \left.D_{b c}\right|_6 \nabla_a u^c + 2 \left.D_{a c}\right|_6 \nabla_b u^c\right) 
       - 2 \left(\left.D_{c b d a}\right|_6 g^{c d} u^b + \left.D_{b d c}\right|_6 g^{b c} \nabla_a u^d \right. \right. \right. \cr
       & \left. \left. \left. - \left.D_{b a c}\right|_6 \nabla^c u^b\right)\right] \xi^a 
      - \left[2 s^b \nabla_{[b} \Theta_{a]} + \Gamma_\s \Theta_a + \left.R_a\right|_1 + \left.D_{a b}\right|_2 \bu^b 
      + \left.R_b\right|_3 \nabla_a u^b + 2 \left.D_{b a c}\right|_5 g^{b c} \right. \right. \cr 
      & \left. \left. + 2 \left.D_{c b d a}\right|_6 g^{c d} u^b + 2 \left.D_{b d c}\right|_6 g^{b c} \nabla_a u^d 
      - 2 \left.D_{b a c}\right|_6 \nabla^c u^b + \nabla^b \left(\left.R_a\right|_3 u_b 
      - \bar{\perp}^c_a \left.R_c\right|_2 \bu_b + 2 \left.D_{a b}\right|_4\right)\right] \bar{\xi}^a \right. \cr
      & \left. + \frac{1}{2} \left[\Psi g^{a b} + \mu^a n^b + \Theta^a s^b + 2 \left.D^{a b}\right|_4 + 2 \left.D^{a b}\right|_5 - \bu^c \left.R_c\right|_2 
      \bu^a \bu^b - u^c \left.R_c\right|_3 u^a u^b \right. \right. \cr
      & \left. \left. + 2 \left.D_{c d}\right|_6 \left(u^a u^b \nabla^{(c} u^{d)} - 2 g^{f a} \perp^{(c}_f \perp^{d)}_e \nabla^b u^e\right) 
      - 2 \nabla_c \left(u^c \left.D^{a b}\right|_6\right)\right] \delta g_{a b}\right\} + {\cal BT} \ ,
\end{align}
where
\begin{subequations}
\begin{align}
         \left.D_{a b}\right|_6 & = \frac{1}{3!} \Theta^{\Ab \Bb \Cb} \frac{\partial s_{\Ab \Bb \Cb}}{\partial \dgde} \pda \peb 
         = \left.D_{b a}\right|_6 \ , \quad 
         u^b \left.D_{a b}\right|_6 = u^b \left.D_{b a}\right|_6 = 0 \label{diss10a} \ , \\ 
         \left.D_{a c b}\right|_6 & = \frac{1}{3!} \Theta^{\Ab \Bb \Cb} \frac{\partial s_{\Ab \Bb \Cb}}{\partial \dgde} \pda \nabla_c \peb = 
         \left.D_{a b c}\right|_6 \ , \quad 
         u^a \left.D_{a c b}\right|_6 = 0 \ , \cr
        & u^c \left.D_{a c b}\right|_6 = - \left.D_{a c}\right|_6 \nabla_b u^c \label{diss10b} \ , \\ 
         \left.D_{a c b d}\right|_6 & = \frac{1}{3!} \Theta^{\Ab \Bb \Cb} \frac{\partial s_{\Ab \Bb \Cb}}{\partial \dgde} 
         \pda \nabla_c \left(\nabla_b \ped\right) = \left.D_{a c d b}\right|_6 \ , \quad 
         u^a \left.D_{a c b d}\right|_6 = 0 \label{diss10c} \ .                
\end{align}
\end{subequations}
The equations of motion are
\begin{subequations}
\begin{align}
2 n^b \nabla_{[b} \mu_{a]} & = \left.R_a\right|_1 + \left.D_{a b}\right|_2 \bu^b + \left.R_b\right|_3 \nabla_a u^b +
      2 \left.D_{b a c}\right|_5 g^{b c}  \cr
       & + \nabla^b \left[\perp^c_a \left.R_c\right|_3 u_b - \left.R_a\right|_2 \bu_b - 2 \left.D_{a b}\right|_5 
       - 2 \left.D_{c d}\right|_6 u_a u_b \nabla^{(c} u^{d)} \right. \cr
       & \left. + 2 \nabla_c \left(u^c \left.D_{a b}\right|_6\right) + 2 \left.D_{c d}\right|_6 \left(\perp^{(c}_b \perp^{d)}_e \nabla_a u^e 
       + \perp^{(c}_a \perp^{d)}_e \nabla_b u^e\right)\right] \cr
       & + 2 \left(\left.D_{c b d a}\right|_6 g^{c d} u^b + \left.D_{b d c}\right|_6 g^{b c} \nabla_a u^d 
       - \left.D_{b a c}\right|_6 \nabla^c u^b\right) \ , \label{peomLg} \\
2 s^b \nabla_{[b} \Theta_{a]} + \Gamma_\s \Theta_a & = - \left.R_a\right|_1 - \left.D_{a b}\right|_2 \bu^b 
       - \left.R_b\right|_3 \nabla_a u^b - 2 \left.D_{b a c}\right|_5 g^{b c} \cr
       & + \nabla^b \left(\bar{\perp}^c_a \left.R_c\right|_2 \bu_b - \left.R_a\right|_3 u_b - 2 \left.D_{a b}\right|_4\right) \cr
       & - 2 \left(\left.D_{c b d a}\right|_6 g^{c d} u^b + \left.D_{b d c}\right|_6 g^{b c} \nabla_a u^d - 
        \left.D_{b a c}\right|_6 \nabla^c u^b\right) \ ,
\end{align}
\end{subequations}
and the energy-momentum-stress tensor is
\begin{align}
   T^{a b} & = \Psi g^{a b} + \mu^a n^b + \Theta^a s^b + 2 \left.D^{a b}\right|_4 + 2 \left.D^{a b}\right|_5 - \bu^c \left.R_c\right|_2 \bu^a \bu^b 
                      - u^c \left.R_c\right|_3 u^a u^b \cr 
                      & + 2 \left.D_{c d}\right|_6 \left[u^a u^b \nabla^{(c} u^{d)} - 2 g^{f a} \perp^{(c}_f \perp^{d)}_e \nabla^b u^e\right] 
                      - 2 \nabla_c \left(u^c \left.D^{a b}\right|_6\right) \ ,
\end{align}
with the generalized pressure $\Psi$, again, taking the same form as before. The entropy creation rate is
\begin{align}
T_g \Gamma_\s & = \left.R_a\right|_1 \bu^a + \left.R_a\right|_2 \bar{a}^a + \left.R_a\right|_3 \left(\bu^b \nabla_b u^a - u^b \nabla_b \bu^a\right) 
          + \left.D_{a b}\right|_2 \bu^a \bu^b - 2 \left.D_{a b}\right|_4 \nabla^{(a} \bu^{b)} \cr
       & + 2 \left.D_{b a c}\right|_5 g^{b c} \bu^a + 2 \left.D_{b a c}\right|_6 \left(g^{b c} \bu^d \nabla_d u^a  
           - \bu^a \nabla^c u^b\right) + 2 \left.D_{c b d a}\right|_6 g^{c d} \bu^a u^b  \ .
\end{align}

As discussed  in Sec.~\ref{fldeqs}, Eq.~\eqref{peomLg} should have only three independent components of motion. It can be shown that 
this is indeed the case by projecting it along $u^a$: the left hand side vanishes identically, and remarkably so does the right hand side. 
Showing the latter boils down to a very tedious exercise in index ``gymnastics'', using the symmetries and various identities of the dissipation 
tensors ``$\left.D_{a b \dots d}\right|_I$'' 
(\eg.~$\left.D_{a b}\right|_6 \nabla^b u^a = \left.D_{b a}\right|_6 \nabla^b u^a = \left.D_{a b}\right|_6 \nabla^a u^b$) and the identities for the 
resistivities ``$\left.R_a\right|_I$''.

From the definition in Eq.~\eqref{diss10a}, we see that $\left.D_{a b}\right|_6$ has only purely spatial pieces with respect to the worldlines of the 
field $u^a$:
\begin{align}
       \left.\tilde{D}_{a b}\right|_6 & = \perp^c_a \perp^d_b \left.D_{c d}\right|_6 = \left.D_{a b}\right|_6 \ ,                  
\end{align}
To get at the modified Cattaneo equation we will again assume the decompositions of $\left.\tilde{R}_a\right|_2$ and $\left.\tilde{R}_a\right|_3$ 
given in Eqs.~\eqref{r2decomp} and \eqref{r3decomp}.

Following the same procedure as before, we project the entropy equation of motion with $u^a$ to find that the entropy creation rate is
\begin{align}
     T^* \Gamma_\s & = - \tilde{s}^a \left\{2 \Bs u^c \nabla_{[c} \tilde{s}_{a]} + \nabla_a T^* 
     + \left(T^* - \frac{s^*}{s^3} \left.r_1\right|_2 \left|\tilde{s}\right|^2\right) a_a \right. \cr
    & \left. + \left(\dot{\cal B}^\s + \nabla^c \left[\frac{1}{s^*} \left(\left.r_1\right|_3 
    - \left(\frac{s^*}{s}\right)^3 \left.r_1\right|_2\right) u_c - \frac{s^*}{s^3} \left.r_1\right|_2 \tilde{s}_c\right]\right) \tilde{s}_a \right. \cr
    & \left. - \frac{2 s^*}{s^3} \left.r_1\right|_2 \tilde{s}^c \nabla_{(c} \tilde{s}_{a)} - \frac{\left|\tilde{s}\right|^2}{s^3} \left.r_1\right|_2 
    \tilde{s}^c \nabla_{(c} u_{a)}\right\} \cr 
    & - 2 \left(\left.\tilde{D}_{a b}\right|_4 + \left.\tilde{D}_{a b}\right|_5\right) \nabla^{(a} u^{b)} 
    - 2 \nabla^b \left(\left.\tilde{D}\right|_4 u_b + \left.\tilde{D}_b\right|_4\right) \cr
     & - 2 \left.\tilde{D}_{a b}\right|_6 u^c \nabla_c \left(\nabla^{(a} u^{b)}\right) 
     + 4 \left.\tilde{D}_{a b}\right|_6 \left(\nabla^{(a} u^{c)}\right) \nabla_c u^b \ . 
     \label{6thentcre}
\end{align}
This has the same structure as in Eq.~\eqref{4thentcre} with the resistivities being confined to the piece which generates the Cattaneo 
equation and the dissipation tensors separated out.

We can now  infer the modified Cattaneo equation from Eq.~\eqref{6thentcre}. After the system has evolved over a number of damping time scales 
$\tau_3$ (which can be inferred as in the previous models), we again expect the entropy drift $\tilde{s}^a$ to be driven to zero, which in turn will force the resistivities to vanish. We are left, again, 
with two field equations for a single fluid model. These can be written
\begin{subequations}
\begin{align}
     2 n^b \nabla_{[b} \mu_{a]} & + 2 s^b \nabla_{[b} \Theta_{a]} + \Gamma_\s \Theta_a  = - 2 \nabla^b \left[\left.\tilde{D}_{a b}\right|_4 
             + \left.\tilde{D}_{a b}\right|_5 + \left.\tilde{D}_{c d}\right|_6 \nabla^{(c} u^{d)} u_a u_b \right. \cr
       & \left. - \nabla_c \left(u^c \left.\tilde{D}_{a b}\right|_6\right) - \left(\left.\tilde{D}_{b c}\right|_6 \nabla_a u^c 
       + \left.\tilde{D}_{a c}\right|_6 \nabla_b u^c\right)\right] \ , \\
    s \left(T_g a_a \right. & \left. + \perp^b_a \nabla_b T_g\right) = - 2 \perp^b_a \left(\left.D_{c b d}\right|_5 g^{c d} 
       + \nabla^c \left.\tilde{D}_{b c}\right|_4  + \left.D_{c d e b}\right|_6 g^{c e} u^d \right. \cr
       & \left. + \left.D_{c d e}\right|_6 g^{c e} \nabla_b u^d \ - \left.D_{c b d}\right|_6 \nabla^d u^c\right) \ .
\end{align}
\end{subequations}
Recall from Eq.~\eqref{dab4decomp} that $\left.\tilde{D}\right|_4 \sim \left|\tilde{s}\right|^2$ and 
$\left.\tilde{D}_b\right|_4 \sim \left|\tilde{s}\right|$ which implies that both vanish in the single fluid limit. Setting the entropy drift and resistivities 
to zero in Eq.~\eqref{6thentcre} leads to
\begin{align}
     T_g \Gamma_\s & = - 2 \left(\left.\tilde{D}_{a b}\right|_4 + \left.\tilde{D}_{a b}\right|_5\right) \nabla^{(a} u^{b)} 
      - 2 \left.\tilde{D}_{a b}\right|_6 u^c \nabla_c \left(\nabla^{(a} u^{b)}\right) 
     \cr
     & + 4 \left.\tilde{D}_{a b}\right|_6 \left(\nabla^{(a} u^{c)}\right) \nabla_c u^b \ . 
\end{align}
Clearly, this is not in an obvious positive-definite form, although it can be made so. However, we believe that it is best to hold off on this step until the next 
section where useful expansions are introduced, using built-in length scales,  through which contact with commonly used equations can be made.

\section{The Entropy ``Drift'' Limit
}
\label{expan}

\noindent
Having provided three sample models for a dissipative two-fluid, we will now try to make contact with the commonly adopted ``single fluid'' perspective. In order to do this, we rewrite the dissipative equations in such a way that the system is explicitly driven to a single-fluid form. In essence, we assume that the system starts off near a state where the entropy drift is zero, the resistive dissipation is 
zero, but the dissipation tensor terms still operate. We then manipulate the equations of motion by assuming the entropy drift is small, 
i.e.,~$\left|\tilde{w}^a\right| \equiv \tilde{s}^a/s^* < 1$, and then expand to linear order in the drift, ${\cal O}\left(\left|\tilde{w}^a\right|\right)$. 

We also appproximate the behaviour by assuming a specific form for $\left.\tilde{D}_{a b}\right|_6$. From the specific Lagrangian written down in 
\cite{andersson24:_ent1diss} (Sec.~VI), we know that a lowest-order form in the Lagrangian (such as $\bginvab \dgab$) for 
$\left.\tilde{D}_{a b}\right|_6$ leads to $\left.\tilde{D}_{a b}\right|_6 \propto h_{a b}$. Given the degrees of freedom included in this model, 
i.e.,~$s_{\Ab \Bb \Cb} =  s_{\Ab \Bb \Cb} \left({\cal S}_X,{\cal S}_V,{\cal S}_g,{\cal S}_{Lg}\right)$, we can reasonably assume that a simple 
model will allow $\left.\tilde{D}_{a b}\right|_6$ to be written as  
\begin{align}
     \left.\tilde{D}_{a b}\right|_6 = \left.\upsilon\right|_6 \left.\tilde{D}\right|_4 \perp_{a b} \ ,
\end{align} 
where the role of $\left.\upsilon\right|_6$ is to ensure consistent units. Obviously this vanishes when the entropy drift vanishes.


Next, we expand the four-momenta $\mu_a$ and $\Theta_a$, making use of the linear drift argument from earlier. From Eq.~\eqref{partmom}, we find
\begin{align}
    \mu_a & = \Bn n u_a + \Asn s \bu_a \approx \Bn n u_a + \Asn s \left(u_a + \tilde{w}_a\right) 
                   = \left(\Bn n + \Asn s\right) u_a + \Asn s \tilde{w}_a \ . 
\end{align}
From its definition $\mu_g = - u^a \mu_a$, we see
\begin{align}
    \mu_g & \approx - u^a \left[\left(\Bn n + \Asn s\right) u_a + \Asn s \tilde{w}_a\right] = \Bn n + \Asn s \quad \Longrightarrow \quad
    \mu_a = \mu_g u_a + \Asn s \tilde{w}_a \ .
\end{align}
From Eq.~\eqref{entmom}, we see
\begin{align}
    \Theta_a & = \Bs s \bu_a + \Asn n u_a \approx \Bs s \left(u_a + \tilde{w}_a\right) + \Asn n u_a  
                   = \left(\Bs s + \Asn n\right) u_a + \Bs s \tilde{w}_a \ . 
                   \label{drftentmom}
\end{align}
From its definition $T_g = - \bu^a \Theta_a$, we see
\begin{align}
    T_g & \approx - \bu^a \left[\left(\Bs s + \Asn n\right) u^a + \Bs s \tilde{w}^a\right] = \Bs s + \Asn n \quad \Longrightarrow \quad
    \Theta_a = T_g u_a + \Bs s \tilde{w}^a \ .
    \label{form_temp}
\end{align}

The approximate equations of motion, energy-momentum-stress tensor, and entropy creation rate for the model of the previous section now become
\begin{subequations}
\begin{align}
2 n^b \nabla_{[b} \left(\mu_g u_{a]}\right) & + 2 n^b \nabla_{[b} \left(\Asn s \tilde{w}_{a]}\right) = s \left.r_1\right|_1 \tilde{w}_a 
          + \frac{s \left.r_1\right|_3}{1 - \left.r_2\right|_3} \tilde{w}_b \nabla_a u^b \cr
       & + \nabla^b \left[s \left(\frac{\left.r_1\right|_3}{1 - \left.r_2\right|_3} - \frac{\left.r_1\right|_2}{1 - \left.r_2\right|_2}\right) \tilde{w}_a u_b 
       - 2 \left.\tilde{D}_{a b}\right|_5 - 2 \left.\tilde{D}_{c d}\right|_6 \nabla^{(c} u^{d)} u_a u_b \right. \cr
       & \left. + 2 \nabla_c \left(u^c \left.\tilde{D}_{a b}\right|_6\right) + 2 \left(\left.\tilde{D}_{b c}\right|_6 \nabla_a u^c 
       + \left.\tilde{D}_{a c}\right|_6 \nabla_b u^c\right)\right] \ , \label{peomLg_exp} \\
2 s^b \nabla_{[b} \left(T_g u_{a]}\right) & + 2 s \tilde{w}^b \nabla_{[b} \left(T_g u_{a]}\right) + 2 s u^b \nabla_{[b} \left(\Bs s \tilde{w}_{a]}\right) 
        + \Gamma_\s T_g u_a = - s \left.r_1\right|_1 \tilde{w}_a \cr
        & - \frac{s \left.r_1\right|_3}{1 - \left.r_2\right|_3} \tilde{w}_b \nabla_a u^b + \nabla^b \left[s \left(\frac{\left.r_1\right|_2}{1 - \left.r_2\right|_2} 
        - \frac{\left.r_1\right|_3}{1 - \left.r_2\right|_3}\right) \tilde{w}_a u_b \right. \cr 
        & \left. - 2 \left(2 \left.\tilde{D}_{(a}\right|_4 u_{b)} + \left.\tilde{D}_{a b}\right|_4\right)\right] - 2 \left.D_{b a c}\right|_5 g^{b c} \ ,
        \label{entLg_exp}
\end{align}
\begin{align}
   T^{a b} & = \Psi g^{a b} + \left(\mu_g n + T_g s + 2 \left.\tilde{D}_{c d}\right|_6 \nabla^{(c} u^{d)}\right) u^a u^b \cr
                     & + \left(T_g s \tilde{w}^a + 2 \left.\tilde{D}^a\right|_4\right) u^b + \left(T_g s \tilde{w}^b + 2 \left.\tilde{D}^b\right|_4\right) u^a \cr
                     & + 2 \left(\left.\tilde{D}^{a b}\right|_4 + \left.\tilde{D}^{a b}\right|_5\right) - 4 g^{c a} \left.\tilde{D}_{c d}\right|_6 \nabla^b u^d 
                      - 2 \nabla_c \left(u^c \left.\tilde{D}^{a b}\right|_6\right) \ ,
\end{align}
and
\begin{align}
     T^* \Gamma_\s & = - \tilde{s}^a \left\{2 \Bs u^c \nabla_{[c} \tilde{s}_{a]} + \nabla_a T^* + T^* a_a 
     + \left(\dot{\cal B}^\s + \nabla^c \left[\frac{1}{s} \left(\left.r_1\right|_3 - \left.r_1\right|_2\right) u_c\right]\right) \tilde{s}_a\right\} \cr 
    & - 2 \left(\left.\tilde{D}_{a b}\right|_4 + \left.\tilde{D}_{a b}\right|_5 - 2 \left.\upsilon\right|_6 \left.\tilde{D}\right|_4 \nabla_{(a} u_{b)}\right) 
    \nabla^{(a} u^{b)} \cr
    & - 2 \left[\left.\dot{\tilde{D}}\right|_4 + \left(\Theta + \left.\upsilon\right|_6 \dot{\Theta} - \left.\upsilon\right|_6 a^a a_a\right) 
    \left.\tilde{D}\right|_4 + \nabla^b \left.\tilde{D}_b\right|_4\right] \  . 
    \label{6thentcre_exp}
\end{align}  
\end{subequations}

Let us now consider the limit in which the entropy drift vanishes, with the consequence of $\left.\tilde{D}\right|_4 \ , \left.\tilde{D}_a\right|_4 \to 0$. 
Again, from the specific Lagrangians written down in \cite{andersson24:_ent1diss} (Sec.~VI), we know that lowest-order forms for 
$\left.\tilde{D}_{a b}\right|_4$ and $\left.\tilde{D}_{a b}\right|_5$ are 
\begin{subequations}
\begin{align}
      \left.\tilde{D}_{a b}\right|_4 = - \left.\alpha\right|_4 \sigma_{a b} - \frac{1}{3} \Theta \left.\beta\right|_4 \perp_{a b} \ , \\
      \left.\tilde{D}_{a b}\right|_5 = - \left.\alpha\right|_5 \sigma_{a b} - \frac{1}{3} \Theta \left.\beta\right|_5 \perp_{a b} \ . 
\end{align}  
\end{subequations}
Similarly, a final lowest order form is 
\begin{align}
    \left.D_{c b d}\right|_5 g^{c d} = u_b \left.\tilde{D}_{c d}\right|_5 \nabla^{(c} u^{d)} \ . 
\end{align} 
Now, we add together Eqs.~\eqref{peomLg_exp} and \eqref{entLg_exp} to find
\begin{align}
\perp^b_a \nabla_b \Psi + \left(\Psi - \Lambda\right) a_a & = - 2 \nabla^b \left(\left.\tilde{D}_{a b}\right|_4 + \left.\tilde{D}_{a b}\right|_5\right)  
          + 2 u_a \left(\left.\tilde{D}_{b c}\right|_4 + \left.\tilde{D}_{b c}\right|_5\right) \nabla^{(b} u^{c)} \ .
          \label{singflevol}
 \end{align} 
The entropy constraint equation becomes, after using the temperature form of the equation in Eq.~\eqref{entLg_exp} and projecting with 
$\perp^b_a$,
\begin{align}
     \perp^b_a \nabla_b \ln T_g + a_a & = \frac{2}{s T_g} \perp_a^b \nabla^c \left(\left.\alpha\right|_4 \sigma_{b c} + \frac{1}{3} 
                               \left.\beta\right|_4 \Theta \perp_{b c}\right) \ .
     \label{Tcons}
\end{align} 
(We consider the Tolman relation for this model in Appendix \ref{appen3}.) This implies that the entropy creation rate limits to
\begin{align}
     T_g \Gamma_\s & = 2 \left(\left.\alpha\right|_4 + \left.\alpha\right|_5\right) \sigma_{a b} \sigma^{a b}
    + \frac{2}{3} \left(\left.\beta\right|_4 + \left.\beta\right|_5\right) \Theta^2 \ , 
    \label{finent}
\end{align} 
which is positive-definite provided the coefficients $\{\left.\alpha\right|_4,\left.\alpha\right|_5,\left.\beta\right|_4,\left.\beta\right|_5\}$ are each 
positive. That is, the model limits to a form that eventually satisfies the second law of thermodynamics. Moreover, the fact that the slow-drift/single-fluid limit leads to a model with the two dissipation channels (shear- and bulk viscosity) anticipated from the standard Navier-Stokes equations is an important proof-of-principle result.

In summary, we have shown that a single-fluid model can be constructed, which has an evolution equation, a constraint equation on the temperature, 
and a positive-definite form for the entropy creation rate. There are four independent coefficients 
$\{\left.\alpha\right|_4,\left.\alpha\right|_5,\left.\beta\right|_4,\left.\beta\right|_5\}$. In Eq.~\eqref{singflevol}, the dissipative tensors add so that 
only two parameters are left; the same is true for the entropy creation rate. It is only the constraint \eqref{Tcons} which breaks the degeneracy ($4 \to 2$) in 
the number of independent parameters.

\section{Concluding remarks}
\label{conclu}

\noindent
We have explored an action principle for relativistic two-fluid systems with dissipation. As proof of principle, we focused on a two-fluid model involving particles and entropy. In order to keep the discussion tractable,  the particle flux creation rate was taken to be zero, while the entropy creation rate was non-zero in accordance with the second law of thermodynamics. The action principle then produces nonlinear evolution equations whose solutions represent dissipative flux vector fields (with non-zero divergences) embedded in curved spacetime solutions to the Einstein field equations. Our extended version of the action includes previously ignored terms---loosely interpreted as velocities---which need to be considered in order to produce equations of motion of the anticipated form, e.g., including terms associated with bulk and shear viscosity. 

The model we have presented recovers known relativistic formulations of the Cattaneo equation, associated with causal heat propagation. Moreover, in the single-fluid limit---where the entropy four-velocity  is locked to that of the matter component---the model reduces to a single field equation alongside a dynamical constraint equation. This constraint represents an extension of the well-known Tolman red-shift condition. The three sample  actions  we considered further demonstrate that the model is able to reproduce (in the single-fluid limit) the expected form of the relativistic Navier-Stokes equations. In general, the results show that the action-based model accords with the expectations in the single fluid limit.
Of course, the general, fully nonlinear, model allows for a much richer structure, the relevance of which remains to be explored and understood. This is a far from straightforward exercise, but efforts in this direction may be required if we want to shed light on dissipative relativistic fluid systems well away from equilibrium.

\vspace{6pt} 


\begin{acknowledgments}

{NA acknowledges support from STFC in the UK via grant number ST/V000551/1. 
TC is supported through the Spanish program Unidad de Excelencia María de Maeztu CEX2020-001058-M financed by MCIN/AEI/10.13039/501100011033 and by the MaX-CSIC Excellence Award MaX4-SOMMA-ICE.}

\end{acknowledgments}

\appendix

\section[\appendixname~\thesection]{More Lagrangian Variations of the Matter Space Metric Derivatives \texorpdfstring{${\cal S}_{Lg}$}
{Matter Space Metric Derivatives}}
\label{appen}
\noindent
As shown in the main body of the text, the starting point for constructing $\bar{\Delta}$ variations of ${\cal S}_{Lg}$ is to use the product 
rule, leading to
\begin{subequations}
\begin{align}
   \bar{\Delta} \dbubgab & = \left(\nabla_a \bgab\right) \bar{\Delta} \bu^a + \bu^a \bar{\Delta} \left(\nabla_a \bgab\right) \ , \\
   \bar{\Delta} \dbungab & = \left(\nabla_a \bgab\right) \bar{\Delta} u^a + u^a \bar{\Delta} \left(\nabla_a \bgab\right) \ , \\ 
   \bar{\Delta} \dbugab & = \left(\nabla_a \gab\right) \bar{\Delta} \bu^a + \bu^a \bar{\Delta} \left(\nabla_a \gab\right) \ , \\
   \bar{\Delta} \dhubgab & = \left(\nabla_a \hgab\right) \bar{\Delta} \bu^a + \bu^a \bar{\Delta} \left(\nabla_a \hgab\right) \ , \\  
   \bar{\Delta} \hdgab & = \left(\nabla_a \hgab\right) \bar{\Delta} u^a + u^a \bar{\Delta} \left(\nabla_a \hgab\right) \ .
\end{align}
\end{subequations}
Following the steps used in the main text we find 
\begin{subequations}
\begin{align}
   \bar{\Delta} \dbubgab & = \bpaa \bpbb \left\{\bu^c \nabla_c \bar{\Delta} g^{a b} + \left[\bu_c \bu_d \nabla^{(a} \bu^{b)}
               - 2 \bar{\perp}^{(a}_c \bar{\perp}^{b)}_e \nabla_d \bu^e\right] \bar{\Delta} g^{c d}\right\} \ , \\
  \bar{\Delta} \bdgab & = \bpaa \bpbb u^c \nabla_c \bar{\Delta} g^{a b} + \left(\bpaa \nabla_c \bpbb\right) \left\{u^c \left[\bar{\Delta} g^{a b} 
                - \frac{1}{2} g^{a b} u_d u_e \Delta g^{d e}\right] \right. \cr
               & \left. + g^{a b} \left[\left(\bar{\xi}^d - \xi^d\right) \nabla_d u^c - u^d \nabla_d \left(\bar{\xi}^c - \xi^c\right)\right]\right\}  
               + \left(\bpbb \nabla_c \bpaa\right) \cr
               & \left\{u^c \left[\bar{\Delta} g^{a b} - \frac{1}{2} g^{a b} u_d u_e \Delta g^{d e}\right]  
               + g^{a b} \left[\left(\bar{\xi}^d - \xi^d\right) \nabla_d u^c - u^d \nabla_d \left(\bar{\xi}^c - \xi^c\right)\right]\right\} \ , \\ 
  \bar{\Delta} \dbugab & = \paa \pbb \bu^c \nabla_c \Delta g^{a b} + \left(\paa \nabla_c \pbb\right) \left[\bu^c \left(\Delta g^{a b} 
                - \frac{1}{2} g^{a b} \bu_d \bu_e \bar{\Delta} g^{d e}\right) \right. \cr
               & \left. + g^{a b} \bu^d \nabla_d \left(\bar{\xi}^c - \xi^c\right)\right]  
               + \left(\pbb \nabla_c \paa\right) \left[\bu^c \left(\Delta g^{a b} - \frac{1}{2} g^{a b} \bu_d \bu_e \bar{\Delta} g^{d e}\right) \right. \cr
               & \left. + g^{a b} \bu^d \nabla_d \left(\bar{\xi}^c - \xi^c\right)\right] + \left(\nabla_c \paa\right) \left(\nabla_d \pbb\right) g^{a b} 
               \left[\bu^c \left(\bar{\xi}^d - \xi^d\right) + \bu^d \left(\bar{\xi}^c - \xi^c\right)\right] \cr
               & + \paa \nabla_c \left(\nabla_d \pbb\right) g^{a b} \bu^c \left(\bar{\xi}^d - \xi^d\right) 
               + \pbb \nabla_d \left(\nabla_c \paa\right) g^{a b} \bu^d \left(\bar{\xi}^c - \xi^c\right) \ , \\  
  \bar{\Delta} \dhubgab & = \paa \bpbb \left\{\bu^c \nabla_c \bar{\Delta} g^{a b} + \frac{1}{2} \left(\nabla^a \bu^b\right) \bu_c \bu_d 
    \bar{\Delta} g^{c d} - \left(\nabla_c \bu^b\right) \left[\bar{\Delta} g^{a c} + \nabla^c \left(\bar{\xi}^a - \xi^a\right) \right] \right. \cr
    & \left. + \bu^c \nabla_c  \left[\nabla^b \left(\bar{\xi}^a - \xi^a\right)\right]\right\} + \left(\bpbb \nabla_c \paa\right) 
    \left[g^{b c} \bu^d \nabla_d \left(\bar{\xi}^a - \xi^a\right) \right. \cr
    & \left.+ \bu^a \left(\bar{\Delta} g^{b c} - \frac{1}{2} g^{b c} \bu_d \bu_e \bar{\Delta} g^{d e}\right) 
    - \left(\nabla^c \bu^b\right) \left(\bar{\xi}^a - \xi^a\right) + \bu^c \nabla^b \left(\bar{\xi}^a - \xi^a\right)\right] \cr
    & + \bpbb \nabla_d \left(\nabla_c \paa\right) g^{a b} \bu^d \left(\bar{\xi}^c - \xi^c\right) \ , \\  
  \bar{\Delta} \hdgab & = \paa \bpbb \left\{u^c \nabla_c \bar{\Delta} g^{a b} + \frac{1}{2} u_c u_d \left(\nabla^b u^a\right) \Delta g^{c d}  - 
     \left(\nabla_c u^a\right) \left[\bar{\Delta} g^{b c} + \nabla^b \left(\bar{\xi}^c - \xi^c\right)\right] \right. \cr
    & \left. + u^c \nabla_c \left[\nabla^b \left(\bar{\xi}^a - \xi^a\right)\right]\right\} + \left(\paa \nabla_c \bpbb\right) 
      \left\{u^c \left[\bar{\Delta} g^{a b} - \frac{1}{2} g^{a b} u_d u_e \Delta g^{d e} + \nabla^b \left(\bar{\xi}^a - \xi^a\right)\right] \right. \cr 
    & \left. + g^{a c} \left[\left(\nabla_d u^b\right) \left(\bar{\xi}^d - \xi^d\right) - u^d \nabla_d \left(\bar{\xi}^b - \xi^b\right)\right]\right\} 
     + \left(\bpbb \nabla_c \paa\right) \left\{g^{b c} \left[\left(\nabla_d u^a\right) \left(\bar{\xi}^d - \xi^d\right) \right. \right. \cr
    & \left. \left. - u^d \nabla_d \left(\bar{\xi}^a - \xi^a\right)\right] + g^{a b} u^d \nabla_d \left(\bar{\xi}^c - \xi^c\right)\right\} 
     + \left(\nabla_c \paa\right) \left(\nabla_d \bpbb\right) g^{a b} u^d \left(\bar{\xi}^c - \xi^c\right) \cr
    & + \bpbb \nabla_d \left(\nabla_c \paa\right) g^{a b} u^d \left(\bar{\xi}^c - \xi^c\right) \ .
\label{dellie2}    
\end{align}
\end{subequations}

\section[\appendixname~\thesection]{Tolman Constraint for More Complicated Spacetimes}\label{app:Tolman}

\noindent
We have shown in the main text that the single-fluid limit of a non-dissipative two-fluid system leads to the standard red-shifted formula for static and spherically symmetric spacetimes. 
It is clear, however, that the constraint (reproduced here for convenience)
\begin{equation}\label{eq:tolman_convenience}
    \perp^b_a\nabla_bT_{\rm g} + T_{\rm g} a_a = 0\;,
\end{equation}
contains more information regarding the behaviour of temperature in dynamical spacetimes, because of the presence of the fluid worldline acceleration. 
To better extract this information we note that the equation is purely spacelike with respect to the worldlines of $u^a$. 
This is one reason why we refer to the extra equation as a ``constraint''; there are no derivatives propagating the temperature from one local time-slice to the next. 

Even with this  lack of ``time-evolution'' for the temperature profile we will carry out a line-integral of Eq.~\eqref{eq:tolman_convenience} along a curve, 
parametrized by the variable $\tau$, whose tangent at each point is $T^a = d x^a/ d \tau$. Because the temperature constraint is spacelike 
with respect to $u^a$, it is only the spacelike piece $\tilde{T}^a = \perp^a_b T^b \equiv d \tilde{x}^a/d \tau$ that ultimately survives as the 
curve cuts across the worldlines of $u^a$. Now, we can write
\begin{align}
   d \tau \tilde{T}^a \partial_a T_{\rm g} + T_{\rm g} d \tau \tilde{T}^a a_a & = 0 \quad \Longrightarrow \quad d \tilde{x}^a \partial_a \ln T_{\rm g} + a_a 
   d \tilde{x}^a = 0 \ . 
   \label{tol_int1}
\end{align} 
Therefore, we have
\begin{align}
     \int^{\tilde{x}^a\left(\tau\right)}_{\tilde{x}^a\left(\tau_0\right)} d \tilde{x}^a \partial_a \left(\ln T_{\rm g}\right) 
     = -  \int^{\tilde{x}^a\left(\tau\right)}_{\tilde{x}^a\left(\tau_0\right)} a_a  d \tilde{x}^a \quad \Longrightarrow \quad 
     T_{\rm g}\left[\tilde{x}^a\left(\tau\right)\right] = T_{\rm g}\left[\tilde{x}^a\left(\tau_0\right)\right] 
     e^{- \int^{\tilde{x}^a\left(\tau\right)}_{\tilde{x}^a\left(\tau_0\right)} a_a d \tilde{x}^a} \ . \label{tol_int2}
\end{align} 

It is easy to check that Eq.~\eqref{tol_int2} does lead back  to Eq.~\eqref{tolrs} by setting $\tau = r$ and $T^a = \left(0,1,0,0\right)$ in Eq.~\eqref{tol_int1}.  
Some more recent demonstrations of the original law restricted to static and stationary fluids and particular equations of state (such as radiation) can be found in \cite{buchdahl49:_temperature,ebert73:_carnot,rovelli11:_thermal,green14:_dynamic,xia24:_generaltolman}; however, these studies are all limited to the perfect, single-fluid system. Here, we have clearly taken the position that the natural starting place for a particle and entropy system is a double fluid, and a benefit of that is that the constraint holds for dynamical spacetimes. It is the two-fluid nature of the approach taken here that has opened up a new path of study of the Tolman relation for generic spacetimes.

\subsection[\appendixname~\thesection]{Tolman Relation for Axisymmetric and Stationary Spacetime}
\label{appen2}
\noindent
Since we derived the familiar Tolman relation as a special case of a more general constraint, let us also explore how these relations are generalized to stationary, axisymmetric spacetime for non-dissipative fluid systems.
According to Eq.~\eqref{eq:tolman_convenience}, 
we need to determine the acceleration. 
Fortunately, we will see that a lot of progress can be obtained by requiring only some specific details 
about the spacetime metric and the four-velocity, with no requirement for explicit solutions.

We start from the metric for an axisymmetric and stationary spacetime which can be written in the form, see e.g. \cite{carter70:_commutation,Carter69:_axistat,bonazzola93:_rotbodies}---we also refer to \cite{Andersson_01:_slowrot} where details and further references for creating this type of spacetime are summarized;
\begin{align}
     d s^2 & = - \left(N^2 - \sin^2 \theta K \left[N^{\phi}\right]^2\right) d t^2 - 2 \sin^2 \theta K N^{\phi} d t d \phi + V d r^2 \cr
               & + K \left(d \theta^2 + \sin^2 \theta d \phi^2\right) \ .
\end{align}
This form is made possible by the existence of two Killing vectors; i.e.,,
\begin{align}
     t^a & = \left(1,0,0,0\right) \ , \quad \phi^a = \left(0,0,0,1\right) \;,
\end{align}
so that all of the metric functions depend on only $r$ and $\theta$. The unit four-velocity $u^a$ can also be shown to take the form
\begin{align}
     u^a & = \frac{t^a + \Omega \phi^a}{\sqrt{N^2 - \sin^2 \theta K \left(N^{\phi} - \Omega\right)^2}} \ ,
\end{align}
where $\Omega$ is the constant, angular velocity. With the metric and four-velocity in hand, we can calculate the four-acceleration
\begin{align}
     a_a & = \left(0,\frac{\partial}{\partial r} \ln \sqrt{N^2 - \sin^2 \theta K \left(N^{\phi} - \Omega\right)^2},\frac{\partial}{\partial \theta} \ln \sqrt{N^2 
                  - \sin^2 \theta K \left(N^{\phi} - \Omega\right)^2},0\right) \ .
\end{align}

With this in hand, we follow the same steps as in the previous section. 
We consider a curve with tangent vector $\tilde{T}^a\left(\tau\right)$ for the line integral, and we take advantage of the fact that the spacetime symmetries imply $T_{\rm g} = T_{\rm g}\left(r,\theta\right)$ to find
\begin{align}
     d \tilde{x}^a \partial_a T_{\rm g} & =  \frac{\partial}{\partial r} \ln T_{\rm g} d r + \frac{\partial}{\partial \theta} \ln T_{\rm g} d \theta 
                                                    \
                                                 = d T_{\rm g} \ .
\end{align}
Similarly, for the second term in \cref{eq:tolman_convenience} we find
\begin{align}
     a_a d \tilde{x}^a & = \frac{\partial}{\partial r} \ln \sqrt{N^2 - \sin^2 \theta K \left(N^{\phi} - \Omega\right)^2} d r \cr 
     & + \frac{\partial}{\partial \theta} \ln \sqrt{N^2 - \sin^2 \theta K \left(N^{\phi} - \Omega\right)^2} d \theta \cr
     & = d \ln \sqrt{N^2 - \sin^2 \theta K \left(N^{\phi} - \Omega\right)^2} \ .
\end{align}
Setting $d T_{\rm g} = - a_a\left[\tilde{x}\left(\tau\right)^b\right] d \tilde{x}\left(\tau\right)^a$ and integrating over the $d \tilde{x}\left(\tau\right)^a$ curve from $\left[r\left(\tau_0\right),\theta\left(\tau_0\right)\right]$ to $\left[r\left(\tau\right),\theta\left(\tau\right)\right]$ gives
\begin{align}
     \frac{T_{\rm g}\left[r\left(\tau\right),\theta\left(\tau\right)\right]}{T_{\rm g}\left[r\left(\tau_0\right),\theta\left(\tau_0\right)\right]} & = 
     \frac{N\left[r\left(\tau_0\right),\theta\left(\tau_0\right)\right] \sqrt{1 - \sin^2 \theta\left(\tau_0\right) \frac{K\left[r\left(\tau_0\right),
     \theta\left(\tau_0\right)\right]}{N^2\left[r\left(\tau_0\right),\theta\left(\tau_0\right)\right]} \left(N^{\phi}\left[r\left(\tau_0\right),\theta\left(\tau_0\right)\right] - 
     \Omega\right)^2}}{N\left[r\left(\tau\right),\theta\left(\tau\right)\right] \sqrt{1 - \sin^2 \theta\left(\tau\right) \frac{K\left[r\left(\tau\right),\theta\left(\tau\right)\right]}
     {N^2\left[r\left(\tau\right),\theta\left(\tau\right)\right]} \left(N^{\phi}\left[r\left(\tau\right),\theta\left(\tau\right)\right] - \Omega\right)^2}} \ .
\end{align}
Therefore, starting at some spacetime point $\left[t\left(\tau_0\right),r\left(\tau_0\right),\theta\left(\tau_0\right),\phi\left(\tau_0\right)\right]$ and 
knowing the initial value $T_{\rm g}\left[r\left(\tau_0\right),\theta\left(\tau_0\right)\right]$ and the metric, the value at the end-point 
$T_{\rm g}\left[t\left(\tau\right),r\left(\tau\right),\theta\left(\tau\right),\phi\left(\tau\right)\right]$ can be determined. 
In particular, using the notation typical of static and spherically symmetric spacetimes---for which we set $N^2=e^\nu$---we note the presence of the co-factor
\begin{align}
         \frac{N\left[r\left(\tau_0\right),\theta\left(\tau_0\right)\right]}{N\left[r\left(\tau\right),\theta\left(\tau\right)\right]} & = 
         e^{- \left\{\nu\left[r\left(\tau\right),\theta\left(\tau\right)\right] - \nu\left[r\left(\tau_0\right),\theta\left(\tau_0\right)\right]\right\}/2} \ ,
\end{align}
which shows how the extended Tolman relation contains the usual overall red-shift term further modulated by rotational effects (c.f. Eq.~\eqref{tolrs}). 
We stress that it is the two-fluid nature of the approach adopted here that automatically yields extended Tolman relations relevant for, say, the modelling of isolated spinning neutron stars. 

\subsection[\appendixname~\thesection]{Tolman Relation for a Spacetime with Dissipation}
\label{appen3}
To conclude, we now specify the generalized Tolman relation to the model discussed in Sec.~\ref{expan}. 
We follow the same steps as in all the previous cases, this time specified to the contraction of Eq.~\eqref{Tcons} to find
\begin{align}
     d \tilde{x}^a \partial_a \ln T_{\rm g} + a_a d \tilde{x}^a & = \frac{2}{s T_{\rm g}} \nabla^b \left(\left.\alpha\right|_4 \sigma_{a b} 
                                  + \frac{1}{3} \left.\beta\right|_4 \Theta h_{a b}\right) d \tilde{x}^a \ .
\end{align} 
We then rewrite this as 
\begin{align}
     \int^{\tilde{x}^a\left(\tau\right)}_{\tilde{x}^a\left(\tau_0\right)} d \tilde{x}^a \partial_a \ln T_{\rm g} & = - \int^{\tilde{x}^a\left(\tau\right)}_{\tilde{x}^a\left(\tau_0\right)} 
     \left[a_a - \frac{2}{s T} \nabla^b \left(\left.\alpha\right|_4 \sigma_{a b} + \frac{1}{3} \left.\beta\right|_4 \Theta h_{a b}\right)\right] d \tilde{x}^a 
     \ .
\end{align} 
and, finally, integrate it to get
\begin{align}
T_{\rm g}\left(\tau\right) & = T_{\rm g}\left(\tau_0\right) e^{- \int^{\tilde{x}^a\left(\tau\right)}_{\tilde{x}^a\left(\tau_0\right)} 
\left[a_a - \frac{2}{s T_{\rm g}} \nabla^b \left(\left.\alpha\right|_4 \sigma_{a b} + \frac{1}{3} \left.\beta\right|_4 \Theta h_{a b}\right)\right] d \tilde{x}^a} \ .
\end{align} 
We stress that to obtain a closed result in this case we do not only need the fluid acceleration, we also require the fluid expansion and shear rates as well as the associated bulk- and shear-viscosity coefficients.

\bibliography{refs_glc.bib}

@article{Andersson_01:_slowrot,
   title={Slowly rotating general relativistic superfluid neutron stars},
   volume={18},
   ISSN={1361-6382},
   url={http://dx.doi.org/10.1088/0264-9381/18/6/302},
   DOI={10.1088/0264-9381/18/6/302},
   number={6},
   journal={Classical and Quantum Gravity},
   publisher={IOP Publishing},
   author={Andersson, N and Comer, G L},
   year={2001},
   pages={969–1002} 
}

@article{andersson24:_sinffldiss,
	doi = {10.20944/preprints202406.0082.v1},
	url = {https://doi.org/10.20944/preprints202406.0082.v1},
	year = 2024,
	month = {June},
	publisher = {Preprints},
	author = {Nils Andersson and Thomas Celora and Gregory Comer and Ian Hawke},
	title = {A Field-Theory Approach for Modeling Dissipative Relativistic Fluids},
	journal = {Preprints}
}

@article{Andersson15:_dissfl_act,
      author         = "Andersson, N. and Comer, G. L.",
      title          = "{A covariant action principle for dissipative fluid
                        dynamics: From formalism to fundamental physics}",
      journal        = "Class. Quant. Grav.",
      volume         = "32",
      year           = "2015",
      number         = "7",
      pages          = "075008",
      doi            = "10.1088/0264-9381/32/7/075008",
      eprint         = "1306.3345",
      archivePrefix  = "arXiv",
      primaryClass   = "gr-qc",
      SLACcitation   = "%%CITATION = ARXIV:1306.3345;%%"
}

@ARTICLE{andersson21:_livrev,
       author = {{Andersson}, Nils and {Comer}, Gregory L.},
        title = "{Relativistic fluid dynamics: physics for many different scales}",
      journal = {Living Reviews in Relativity},
         year = 2021,
        month = {Dec.},
       volume = {24},
       number = {1},
          eid = {3},
        pages = {3},
          doi = {10.1007/s41114-021-00031-6},
archivePrefix = {arXiv},
       eprint = {2008.12069},
 primaryClass = {gr-qc},
       adsurl = {https://ui.adsabs.harvard.edu/abs/2021LRR....24....3A}
}

@Article{andersson24:_ent1diss,
AUTHOR = {Andersson, Nils and Celora, Thomas and Comer, Gregory and Hawke, Ian},
TITLE = {A Field-Theory Approach for Modeling Dissipative Relativistic Fluids},
JOURNAL = {Entropy},
VOLUME = {26},
YEAR = {2024},
NUMBER = {8},
ARTICLE-NUMBER = {621},
URL = {https://www.mdpi.com/1099-4300/26/8/621},
ISSN = {1099-4300},
DOI = {10.3390/e26080621}
}

@article{bonazzola93:_rotbodies,
  author   = {Bonazzola, S. and Gourgoulhon, E. and Salgado, M. and Marck, J.-A.},
  format   = {print},
  journal  = {Astron. Astrophys.},
  pages    = {421--443},
  title    = {Axisymmetric rotating relativistic bodies: a new numerical approach for `exact' solutions},
  volume   = {278},
  year     = {1993}
}

@article{buchdahl49:_temperature,
  title={Temperature equilibrium in a stationary gravitational field},
  author={Buchdahl, HA},
  journal={Physical Review},
  volume={76},
  number={3},
  pages={427},
  year={1949},
  publisher={APS}
}

@incollection{carter76,
  address   = {New Delhi, India},
  author    = {Carter, B.},
  booktitle = {Journees Relativistes},
  editor    = {Cahen, M. and Debever, R. and Geheniau, J.},
  format    = {print},
  pages     = {12--27},
  publisher = {Wiley Eastern, New Delhi},
  title     = {The Canonical Treatment of Heat Conduction and Superfluidity in Relativistic Hydrodynamics},
  year      = {1976}
}

@article{Carter69:_axistat,
    author = {Carter, Brandon},
    title = {Killing Horizons and Orthogonally Transitive Groups in Space‐Time},
    journal = {Journal of Mathematical Physics},
    volume = {10},
    number = {1},
    pages = {70-81},
    year = {1969},
    month = {01},
    issn = {0022-2488},
    doi = {10.1063/1.1664763},
    url = {https://doi.org/10.1063/1.1664763}
}

@article{carter70:_commutation,
  author        = {Carter, B.},
  cited         = {14 December 2006},
  doi           = {10.1007/BF01647092},
  format        = {print},
  journal       = {Commun. Math. Phys.},
  onlineversion = {http://projecteuclid.org/getRecord?id=euclid.cmp/1103842335},
  pages         = {233--238},
  title         = {The Commutation Property of a Stationary, Axisymmetric System},
  volume        = {17},
  year          = {1970}
}

@Article{Cattaneo48:_cateq,
author = "Cattaneo, C.", 
year = "1948", 
journal = "Atti Seminario Univ. Modena",
volume = 3,
pages = 83,
title = {On Heat Conduction}
}

@ARTICLE{celora21:_lindiss,
       author = {{Celora}, T. and {Andersson}, N. and {Comer}, G.~L.},
        title = "{Linearizing a non-linear formulation for general relativistic dissipative fluids}",
      journal = {Class. and Quantum Grav.},
         year = 2021,
        month = {Mar.},
       volume = {38},
       number = {6},
          eid = {065009},
        pages = {065009},
          doi = {10.1088/1361-6382/abd7c1},
archivePrefix = {arXiv},
       eprint = {2008.00945},
 primaryClass = {gr-qc},
       adsurl = {https://ui.adsabs.harvard.edu/abs/2021CQGra..38f5009C}
}

@article{comer94:_hamil_sf,
  author   = {Comer, G.L. and Langlois, D.},
  doi      = {10.1088/0264-9381/11/3/021},
  format   = {print},
  journal  = {Class. Quantum Grav.},
  pages    = {709--721},
  title    = {Hamiltonian Formulation for Relativistic Superfluids},
  volume   = {11},
  year     = {1994}
}

@article{comer93:_hamil_multi_con,
  author   = {Comer, G.L. and Langlois, D.},
  doi      = {10.1088/0264-9381/10/11/014},
  format   = {print},
  journal  = {Class. Quantum Grav.},
  pages    = {2317--2327},
  title    = {Hamiltonian Formulation for Multi-constituent Relativistic Perfect Fluids},
  volume   = {10},
  year     = {1993}
}

@article{eckart40:_rel_diss_fluid,
  author   = {Eckart, C.},
  doi      = {10.1103/PhysRev.58.919},
  format   = {print},
  journal  = {Phys. Rev.},
  pages    = {919--924},
  title    = {The Thermodynamics of Irreversible Processes. III. Relativistic Theory of the Simple Fluid},
  volume   = {58},
  year     = {1940}
}

@article{ebert73:_carnot,
  title={Carnot cycles in general relativity},
  author={Ebert, R and G{\"o}bel, R},
  journal={General Relativity and Gravitation},
  volume={4},
  number={5},
  pages={375--386},
  year={1973},
  publisher={Springer}
}

@article{green14:_dynamic,
  title={Dynamic and thermodynamic stability of relativistic, perfect fluid stars},
  author={Green, Stephen R and Schiffrin, Joshua S and Wald, Robert M},
  journal={Classical and Quantum Gravity},
  volume={31},
  number={3},
  pages={035023},
  year={2014},
  publisher={IOP Publishing}
}

@article{Israel79:_kintheo1,
  author   = {Israel, W. and Stewart, J.M.},
  format   = {print},
  journal  = {Ann. Phys. (N.Y.)},
  pages    = {341--372},
  title    = {Transient Relativistic Thermodynamics and Kinetic Theory},
  volume   = {118},
  year     = {1979}
}

@book{landau1980:_statistical,
  title={Statistical Physics, Part 1},
  author={Landau, L.D. and Lifshitz, E.M.},
  year={1980},
  publisher={Pergamon Press},
  edition={3rd},
  series={Course of Theoretical Physics},
  volume={5}
}

@BOOK{Landau69:_mech_book,
       author = {{Landau}, L.~D. and {Lifshitz}, E.~M.},
        title = "{Mechanics}",
         year = 1976,
      edition = {3rd},   
      publisher = {Elsevier},   
  series={Course of Theoretical Physics},
  volume={1}
}

@book{mtw73,
  address   = {San Francisco, U.S.A.},
  author    = {Misner, C.W. and Thorne, K.S. and Wheeler, J.A.},
  format    = {print},
  publisher = {W.H. Freeman},
  title     = {Gravitation},
  year      = {1973}
}

@article{Monsalvo10:_thermogr,
   title={Thermal dynamics in general relativity},
   volume={467},
   ISSN={1471-2946},
   url={http://dx.doi.org/10.1098/rspa.2010.0308},
   DOI={10.1098/rspa.2010.0308},
   number={2127},
   journal={Proceedings of the Royal Society A: Mathematical, Physical and Engineering Sciences},
   publisher={The Royal Society},
   author={Lopez-Monsalvo, C. S. and Andersson, N.},
   year={2010},
   month={Sep}, 
   pages={738–759} 
 }

@misc{monsalvo11:_thesis,
      title={Covariant Thermodynamics and Relativity}, 
      author={C S Lopez-Monsalvo},
      year={2011},
      eprint={1107.1005},
      archivePrefix={arXiv},
      primaryClass={gr-qc},
      url={https://arxiv.org/abs/1107.1005}, 
}

@article{Pound59:_red-shift,
  title = {Gravitational Red-Shift in Nuclear Resonance},
  author = {Pound, R. V. and Rebka, G. A.},
  journal = {Phys. Rev. Lett.},
  volume = {3},
  issue = {9},
  pages = {439--441},
  numpages = {0},
  year = {1959},
  month = {Nov},
  publisher = {American Physical Society},
  doi = {10.1103/PhysRevLett.3.439},
  url = {https://link.aps.org/doi/10.1103/PhysRevLett.3.439}
}

@book{reichl98:_book,
  address   = {Austin, U.S.A.},
  author    = {Reichl, L.E.},
  format    = {print},
  publisher = {University of Texas Press},
  title     = {A Modern Course in Statistical Physics},
  year      = {1984}
}

@article{rovelli11:_thermal,
  title={Thermal time and Tolman--Ehrenfest effect:‘temperature as the speed of time’},
  author={Rovelli, Carlo and Smerlak, Matteo},
  journal={Classical and Quantum Gravity},
  volume={28},
  number={7},
  pages={075007},
  year={2011},
  publisher={IOP Publishing}
}

@article{stew77,
  author   = {Stewart, J.M.},
  format   = {print},
  journal  = {Proc. R. Soc. London, Ser. A},
  pages    = {59--75},
  title    = {On transient relativistic thermodynamics and kinetic theory},
  volume   = {357},
  year     = {1977}
}

@Book{strang80:_lin_alg,
  author    = {G. Strang},
  title     = {Linear Algebra and Its Applications},
  edition   = {2nd},
  publisher = {New York: Academic Press},
  year      = 1980
}

@book{tolman34:_book,
  address   = {New York, U.S.A.},
  author    = {Tolman, R.C.},
  format    = {print},
  note      = {Reprint of 1934 edition},
  publisher = {Dover Publications},
  title     = {Relativity, Thermodynamics, and Cosmology},
  year      = {1987}
}

@article{xia24:_generaltolman,
  title={General proof of the Tolman law},
  author={Xia, Minghao and Gao, Sijie},
  journal={The European Physical Journal Plus},
  volume={139},
  number={6},
  pages={500},
  year={2024},
  publisher={Springer}
}

\end{document}